\documentclass[11pt]{article}
\usepackage[bookmarks,bookmarksnumbered=true,colorlinks=true,
  linkcolor=blue,urlcolor=blue,citecolor=blue,breaklinks = true,backref=page]{hyperref}
\usepackage{mathtools}

\usepackage{color}
\usepackage{url}
\usepackage[nodayofweek]{datetime}
\usepackage{amsmath}
\usepackage{amsthm}
\usepackage[round]{natbib}
\usepackage{graphicx}
\usepackage{booktabs} 
\usepackage{xr-hyper}
\usepackage{scalerel,amssymb}
\usepackage{rotating}

\usepackage{array}
\newcolumntype{L}{>{\centering\arraybackslash}m{5cm}}

\input{commands.sty}

\usepackage[pagewise, displaymath, pagewise]{lineno}

\newcommand{\limitSlope}{\kappa}
\newcommand{\logSLow}{\log\!S_{\Lowx}}

\newcommand{\logDeltaHighLow}{\log\!\Delta_{\SHighx\SLowx}}
\newcommand{\qlogisp}{\textrm{qlogis}p}
\newcommand{\logDeltaSLowE}{\log\!\Delta_{\SLowx E}}
\newcommand{\logDeltaSlopes}{\log\!\Delta_{\textrm{Slopes}}}
\newcommand{\logSigmaX}{\log\!\sigma_{X}}
\newcommand{\Lowx}{\textrm{Low}}
\newcommand{\SLowx}{\textrm{SLow}}
\newcommand{\Highx}{\textrm{High}}
\newcommand{\SHighx}{\textrm{SHigh}}
\newcommand{\Midx}{\textrm{Mid}}

\newcommand{\MidU}{\textrm{Mid-U}}
\newcommand{\MidL}{\textrm{Mid-L}}

\newcommand{\plogis}{\texttt{plogis}}
\newcommand{\qlogis}{\texttt{qlogis}}

\newcommand{\footremember}[2]{%
	\footnote{#2}
	\newcounter{#1}
	\setcounter{#1}{\value{footnote}}%
}

\newdateformat{mydate}{\twodigit{18}{  }March 2024}

\usepackage{amsthm}
\title{Robust Numerical Methods for Nonlinear Regression}
\author{%
  Peng Liu \footremember{jmp}{JMP Statistical Discovery LLC}
  \and
  William Q. Meeker\footremember{isu}{Department of Statistics, Iowa State University}
}
\begin{document}
\mydate

\maketitle

\begin{abstract}
Many scientific and engineering applications require fitting
regression models that are nonlinear in the parameters. Advances in
computer hardware and software in recent decades have made it easier
to fit such models. Relative to fitting regression models that are
linear in the parameters, however, fitting nonlinear regression
models is more complicated.  In particular, software like the
\texttt{nls} R function requires care in how the model is
parameterized and how initial values are chosen for the maximum
likelihood iterations. Often special diagnostics are needed to
detect and suggest approaches for dealing with identifiability
problems that can arise with such model fitting. When using Bayesian
inference, there is the added complication of having to specify
(often noninformative or weakly informative) prior distributions. Generally,
the details for these tasks
must be determined for each new nonlinear regression model. This paper
provides a step-by-step procedure for specifying these details
for any appropriate nonlinear regression model. Following the procedure
will result in a numerically robust algorithm for fitting the nonlinear
regression model. We illustrate the methods with three different nonlinear
models that are used in the analysis of experimental fatigue
data and we include two detailed numerical examples.
\end{abstract}


\newpage
\tableofcontents

\newpage

\section{Introduction}
\label{section:numerical.methods.posterior}
\subsection{Background and motivation}
\label{section:background.and.motivation}

Numerically robust algorithms for estimating the parameters
of nonlinear regression models, using nonlinear least squares (NLS),
maximum likelihood (ML)
or Bayesian estimation, require
\begin{itemize}
\item
Careful attention to parameterization.
\item
Methods for finding ``initial values''
for the parameters to start iterative estimation algorithms.
\item
Diagnostics for detecting estimability problems when the data are
not sufficient to estimate the model parameters (e.g., trying to fit
a four-parameter curvilinear relationship when there is not
sufficient curvature in the data).
\item
When doing Bayesian estimation, default noninformative or weakly
informative priors.
\end{itemize}

Over many years in many application areas,
we have had experiences in which we developed and
used the methods given in this paper to produce robust computational
algorithms across a wide range of different kinds of nonlinear
statistical models. The motivation to write this
paper came from recent work to develop the
robust numerical methods for the nonlinear regression models used in
the applications in \citet{Meekeretal2022}. The methods presented here,
however, are broadly applicable for many other application
areas where nonlinear estimation is used.
Some other examples that have
been documented are described in \citet[][Chapters~10, 20, 21,
  and~22]{MeekerEscobarPascual2021} and
\citet{TianLewis-BeckNiemiMeeker2022}. 

\subsection{Other related literature}
Before computers became widely available to researchers, nonlinear
regression models were rarely used in practice. Starting in the late
1960s, rapid progress began. In pioneering work, \citet{Ross1970}
showed the importance studying the analytical properties of a
nonlinear model and its relationship to the data. Doing so helps to
formulate numerically robust estimation algorithms that have a much higher
probability of successfully completing the estimation task. His
ideas were extended and described, along with many additional
examples in \citet{Ross1990}. The ideas of Ross have had a strong
effect on the methods we have developed and refined.

Other books on nonlinear regression
from around the same time include \citet{Ratkowsky1983},
\citet{Gallant1987}, 
\citet{BatesWatts1988}, and \citet{SeberWild1989}.
More recently,
\citet{HuetBouvierPoursatJolivet2004} use a large number of example
applications to illustrate the use of the \textit{nls2} \Rsoftware{} package to
fit nonlinear regression models.  \citet{Nash2014a} reviews methods
of nonlinear estimation with emphasis on the mathematical
programming aspects of solving NLS or
ML estimation problems.

\subsection{Overview}
The rest of this paper is organized as follows.
Section~\ref{section:general.strategy} outlines the
general strategy for developing a numerically robust estimation procedure.
Section~\ref{section:fatigue.models.introduction} provides a brief
introduction to linear and nonlinear regression models
used to describe experimental fatigue data.
We use nonlinear regression models from
this area to illustrate the application of the
main ideas presented in this paper.
Section~\ref{section:parameterization.initial.values.coffin.manson.model}
provides implementation details for the Coffin--Manson
nonlinear regression model.
Sections~\ref{section:parameterization.initial.values.nishijima.hyperbolic.model}
and~\ref{section:parameterization.initial.values.box.cox.loglinear.model}
do the same for the Nishijima and
Box--Cox/Loglinear-$\sigma_{N}$ models, respectively.
Section~\ref{section:numerical.examples} contains numerical examples
that illustrate the methods.
Section~\ref{section:concluding.remarks} provides some
concluding remarks and suggests areas for related future research.

\section{The General Strategy}
\label{section:general.strategy}
This section outlines our general strategy for developing robust
numerical methods for fitting nonlinear regression models.
Our focus is on models that have a single explanatory variable, but
all of the ideas extend directly to models with more than one
explanatory variable.
Briefly, from many years of experience, we have found that numerically robust
nonlinear estimation algorithms can be developed by using a good
parameterization and having good initial values for the estimation
algorithm. 

As suggested in the previous paragraph, a key component of our
strategy is to carefully choose a parameterization for a statistical
model. Proposed linear and nonlinear regression models are often
written in a form where one or more of the traditional parameters
(hereafter, the TPs) have no practical interpretation and
may have one or more high correlations between parameters;
such parameters can be ``unstable.'' An example is the intercept of the simple
linear regression model that is far removed from the data. One
effect of such a poor parameterization (beside lack of practical
interpretability of the intercept) is that the estimates of the
slope and intercept can be highly correlated (and thus relatively
unstable). The common remedy for this particular deficient parameterization
(a remedy that is also useful in other regression applications)
is to center the explanatory variable which, in effect, redefines
the intercept to be at the center of the data.

Sections~\ref{section:parameterization.initial.values.coffin.manson.model}--\ref{section:parameterization.initial.values.box.cox.loglinear.model}
provide technical details for the suggested scaling and parameterization
for the three example nonlinear regression models.
Section~\ref{section:numerical.examples} provides two numerical
examples that illustrate the ML and data analysis steps of the
procedures described in this section.

\begin{table}[!tbp]
\caption{Acronyms used in this paper}
\label{table:acronyms}
\begin{center}
\begin{center}
\begin{tabular}{lll}
cdf & \quad &Cumulative distribution function\\
pdf & \quad &Probability density function\\
ML& \quad &Maximum likelihood\\
NLS& \quad &Nonlinear least squares\\
TP& \quad &Traditional parameter based on scaled data\\
SP& \quad &Stable parameter\\
USP& \quad &Unrestricted stable parameter\\
TPNS& \quad &Traditional parameter if data had not been scaled\\
\SN{}  & \quad &Stress and number of cycles
\end{tabular}
\end{center}
\end{center}
\end{table}

\subsection{Steps for ML estimation}
\label{section:steps.ml.estimation}
Even when the final goal is Bayesian
inference, we find it useful
(for reasons described in Section~\ref{section:steps.bayesian.estimation})
to start with ML estimation.  We use ML instead of NLS
because ML is more general. It provides a statistically correct method for
handling the  right-censored observations (known as ``runouts'' in
the fatigue literature) and distributions other than the normal
distribution that are commonly seen in
many applications. Thus,
whether doing ML or Bayesian estimation, we start with the following
procedures for ML estimation.

\begin{enumerate}
\item
\label{step:scale.data}
Scale both the response and the explanatory variables.
Dividing the values of each variable by the largest
values of each variable
generally works well.
This will ensure that the performance of
the estimation algorithms will not depend on the units of the
variables and avoids numerical problems that might arise (e.g., with the
Box--Cox model in
Section~\ref{section:parameterization.initial.values.box.cox.loglinear.model})
when large numbers are raised to a large negative
power.
\item
\label{step:stable.parameters}
Identify ``stable parameters'' (SPs) \citep[as defined
  by][]{Ross1970,Ross1990} that will not be highly correlated and that can be
identified from available data. \citet[][in Chapters 3 and
  4]{SeberWild1989} also describe the importance of parameterization
in nonlinear regression.  For example, parameters that can be
identified as features of the data in a plot of the model fitted to data will
often be stable.

Because appropriate SPs are not unique
and can be usefully defined in different ways, finding such SPs
is as much art as science and often requires some
trial-and-error experimentation for a particular model by using a
collection of typical data sets.
One criterion is to choose an SP definition that results in
low estimated correlations obtained from the estimated SPs
variance-covariance matrix.
The choice among different definitions
for SPs usually is not critical but we have found that by
experimenting with a benchmark collection of data sets (some of which might
have been simulated) allows comparison among alternative SP
definitions.

It is appealing to have SPs that are
interpretable (and usually they are, almost by
definition, if they can be identified as features of the data in a plot).
Also, when Bayesian methods are to
be used, SPs will generally simplify prior
specification and elicitation from subject-matter experts
because such parameters typically have an
easy-to-understand interpretation and
have estimates that tend not to be highly correlated.

It is possible for the SPs to depend on the available data. For
example, \citet[][Chapter 10]{MeekerEscobarPascual2021} and
\citet[][Section~3.2]{TianLewis-BeckNiemiMeeker2022} suggest using a
distribution quantile (to replace the usual scale parameter)
as an SP when fitting log-location-scale
distributions to censored data. The particular quantile that should
be used depends on the amount of censoring in the data. In two of the
examples in this paper,
we use points on the fitted model
that are chosen to be \textit{within the range of the data} as SPs.

\item
\label{step:unrestricted.parameters}
If needed, further transform the SPs so that they
do not have limits depending on other parameters and are
unrestricted (e.g., take logs of positive parameters). Most
estimation algorithms (ML or Bayesian MCMC) perform better when
parameters are unrestricted. For clarity and consistency, we call
these parameters the ``unrestricted stable parameters'' (USPs)
although they could also be called the ``estimation
parameters''  because they are the parameters that the
estimation/optimization algorithm sees.
An important consideration in the definition of SPs and USPs is that
there should be an easy way to compute the TPs as a function of
the USPs.

\item
\label{step:initial.values}
Initial values are needed to start ML iterations and MCMC
algorithms.
Recall that SPs can often be
thought of as features that can be identified from plots of the data.
Usually initial values can be obtained by using simple
descriptive statistics or moment
estimates (e.g., sample means, variances, and ordinary least
squares), ignoring any censoring or other complicated features of the
data.
Especially when using a stable parameterization, the initial values
to start ML estimation do not have to be close to the maximum of the
likelihood.
When the likelihood is well behaved (i.e., when the likelihood is
expressed in terms of stable parameters and
when the maximum of the likelihood is unique),
initial values only need to be in the
region where the likelihood is nonnegligible.

In situations where the likelihood has multiple maxima (usually one
global maximum with one or more local maxima at lower levels of
relative likelihood), a more elaborate
approach is needed. For example, it may be necessary to perform the
optimization with many systematically or randomly chosen start
values. Generally, however, it is best to understand the reason and
interpretation of the multiple maxima (e.g., such multiple maxima
may suggest an over-parameterized model). We illustrate this in one of
our examples in Section~\ref{section:estimation.Polynt.data}.
\item
Attempt the ML estimation. Check the results by computing the
gradient vector (the elements should be close to zero) and the
Hessian matrix (the eigenvalues should all be negative) of the
log-likelihood evaluated at the maximum. We have found that
(especially because we work with USPs) the needed derivatives can be
computed by using carefully programmed finite differences \citep[e.g.,][]{Barton1992}.
\item
When the estimation diagnostics indicate
that convergence has \textit{not} been
successful, the particulars (e.g., the eigenvector corresponding to
the smallest eigenvalue of the Hessian matrix) will generally provide signatures
indicating what caused the problem and that, in turn, will suggest
how to proceed (e.g., by fixing a parameter, switching to a
particular submodel, a limiting model,
or some other alternative well-fitting model, perhaps with fewer parameters). 
\item
To check for estimability and as a diagnostic to help
understand the root-causes of estimation problems, especially in
situations with a new model and a new kind of data, it is important to
examine profile relative likelihood plots for individual parameters
(one-dimensional profiles) and parameter pairs (two-dimensional
profiles).
The examples in
Sections~\ref{section:estimation.Inconel.data}
and \ref{section:estimation.Polynt.data}
illustrate the use of such profiles.
\item
Translate the USP ML estimates back to ML estimates of the
TPs that users would have obtained if without reparameterization
(and if there were no numerical problems).
Also obtain the variance-covariance matrix for the
TPs using the vector version of the delta method \citep[e.g.,
  Section C.2 of][]{MeekerEscobarPascual2021}.
\item
  Use the estimation results to do the usual residual analysis to
  assess how well the fitted model agrees with the data.
\item
Translate the TP ML parameter estimates back to ML estimates of the
unscaled TPs (which we call the TPNSs) that users would have obtained
if unscaled data had been used (and if there were no numerical
problems). Also obtain the corresponding
variance-covariance matrix for the TPNSs.
\item
\label{step:tp.for.ml.postprocessing}
In our software, we use the TP estimates (not the TPNS
parameter estimates) to compute point estimates
and confidence intervals for needed functions of the parameters
such as distribution quantiles or cdf tail
probabilities. Using the TP estimates
protects against numerical problems that might occur
when using the TPNSs.  Parameter estimation
results reported to users, however, are always
unscaled (i.e., the TPNSs as if the data had not been scaled). Users of the
software generally have no need to know about the details of how
the scaling was~done.
\item
To develop numerically robust
algorithms for particular models, it is important to employ ``stress
tests'' by collecting, simulating, or otherwise generating 
a benchmark collection of data sets designed to
challenge the algorithms. Testing on such a collection of data sets
helps to develop appropriate diagnostic checks
to trap potential estimation problems (e.g., when the level of a
profile likelihood is non-zero for extreme values
(i.e., $\pm \infty$) of one or more of the USPs).
\end{enumerate}

\subsection{Steps for Bayesian prior specification and estimation}
\label{section:steps.bayesian.estimation}
If continuing on to Bayesian estimation, we again use the
USPs as estimation parameters and suppose that there is a
desire or need to use noninformative
or weakly informative prior distributions. Then we use the following
steps to obtain default prior distributions, MCMC initial values, and
posterior draws via an MCMC engine. In this paper we take the modern
applied approach \citep[also see the rejoinder for the discussion
in][]{TianLewis-BeckNiemiMeeker2022}
and treat the MCMC engine as a black box. The
justification for doing this in our class of problems (models with
at most a moderate number of parameters
and the number of parameters does not depend on the
data) is that with properly-defined USPs and confidence that all
parameters are estimable, a reasonable MCMC  algorithm (e.g.,
programing our USP model in Stan) should complete successfully.

In all cases, it is important to carefully examine MCMC diagnostics
(especially trace plots for three or four independently drawn
chains and corresponding numerical summaries)
and check the adequacy of the fit to the data.
\begin{enumerate}
\item
\label{step:flat.prior}
With typical \SN{} data sets (which contain much information about
the parameters of the \SN{} relationship), there is little or no risk of
encountering an improper posterior distribution. Thus using flat
(i.e., uniform over the entire real line) marginal priors for each
of the USPs
provides a natural default joint prior distribution for the USPs
defined in Step~\ref{step:unrestricted.parameters}
of Section~\ref{section:steps.ml.estimation}.
In many cases, the flat
prior for the USP will provide a good
approximation to a reference prior \citep[e.g.,][]{BergerBernardo1992a}.
A reference prior would be difficult to
establish exactly because of the nonlinearity in some
\SN{} relationships and right censoring typically seen at
low levels of stress.
\item
While the flat prior suggested in the previous step should work well
for most model/data combinations (especially if the data/model
combination provided valid ML estimates), it is possible that the
MCMC sampler could encounter difficulties
(e.g., because the improper prior puts non-zero probability in
nonsensical parts of the parameter space where the log-likelihood
evaluations are in the noise of the numerical computations). In
extreme cases, the posterior could be improper (e.g., when
there is little information in the data about one or more
parameters of the nonlinear regression model).
In such cases, one can replace the flat marginal priors with
approximately flat normal (Gaussian) distributions with a
large standard deviation.
Such an approximately noninformative (or weakly informative) prior will
ensure that the posterior is proper but will not ensure that the
MCMC sampler will accurately sample from the joint posterior
distribution. We have encountered situations where the MCMC sampler
finishes, the MCMC diagnostics look good, but the fitted model does
not agree with the data (because ``posterior draws'' were being
taken from one part of the approximately flat prior, far away from
the likelihood). We have learned, however, that these problems are
largely avoided if the USPs are defined well, as we will discuss
within some of our particular
numerical examples in Section~\ref{section:numerical.examples}.
\item
When specifying a non-flat prior distribution (whether informative
or weakly informative) we recommend using the ``parameter-range''
method
\citep[as used, for example, in Chapter 10 of][]{MeekerEscobarPascual2021}
and \citet[][Section~7.3]{TianLewis-BeckNiemiMeeker2022}
to specify a two-parameter marginal prior distribution
(e.g., a normal distribution or a Student’s~$t$ distribution with
given degrees of freedom). Instead of
specifying a marginal prior distribution and parameters of that distribution,
specify instead the
distribution and the 0.005 and 0.995 quantiles of the distribution
as the ``range'' of the distribution.
Such ranges are generally easier to elicit from subject-matter experts.
\item
\label{step:weakly.informative}
When flat or extremely wide normal distributions do not provide a
satisfactory noninformative prior, it is necessary to use a weakly
informative prior, usually represented by normal distributions with
standard deviations that are not too large. To choose the mean and
standard deviation of those normal distributions one can use
information such as the scale of ones data, previous experience, and
other engineering knowledge.

\item
From our experience, having to specify weakly informative prior distributions
in the manner described in Step~\ref{step:weakly.informative} generally
takes much time because it
requires tedious trial-and-error. Thus (especially in software being designed
to be user friendly) there is a need to
automate the process of finding \textit{numerically-stable} default
noninformative or
minimally informative priors. Technically, one should
not use information from the data to help set the prior distribution
(doing so violates the likelihood principle). For example, it would
be a serious mistake to use 95\% or even 99\% confidence intervals
from the data to choose a prior distribution parameter range
(doing so is like using the available data twice!).
Nevertheless, from a
practical point of view, one can use the results of an ML estimation
in a manner that can be shown to provide a prior distribution that
is effectively noninformative and that avoids substantive reuse of one's data.

For example, if Wald-like approximate confidence
intervals were computed using a factor of 20 standard errors instead
of 1.96, the prior would be approximately flat over the part of the
parameter space and somewhat beyond where the likelihood is
nonnegligible. Weakly informative priors chosen in this way should
be insensitive to the exact values of the mean and standard deviations of the
specified normal distributions and this can be checked by doing
sensitivity analysis (e.g., by generating a few hundred joint
posterior draws and comparing for different settings of the
prior-specification algorithm). For a similar view (but a somewhat
different approach), see the vignette for the \texttt{rstanarm}
\Rsoftware{} package in \citet{GabryGoodrich2020}.
\item
Following common practice, we run four independent chains so that we
can use standard MCMC diagnostics to check for convergence.
We have found that MCMC initial
values obtained by sampling randomly, without replacement, from
the vertices of a hyper-rectangle defined by the endpoints of 95\%
confidence intervals for the parameters works well. Doing this improves the
chances that MCMC sampling will be from the joint posterior instead of
some alternative low-level island of probability elsewhere in the
parameter~space.
\item
After draws from the posterior of the USPs have been
computed and checked, similar to what is suggested in
Step~\ref{step:tp.for.ml.postprocessing} in
Section~\ref{section:steps.ml.estimation},
compute and save posterior draws for the
marginal distributions of the TPs (along with meta information on
how the data were scaled to ensure that final results are reported
correctly in terms what would have been computed by using the
unscaled data).
These can
be used for post processing of the results to compute estimates and
credible intervals for quantities of interest and make corresponding
plots. For reasons of numerical stability
when doing the post processing, it is better to \textit{not} to
convert these draws back to the TPs for the unscaled data (i.e.,
back to the TPNS).
\end{enumerate}

\section{A Brief Introduction to Modern Statistical Models
  for Experimental Fatigue Data}
\label{section:fatigue.models.introduction}
 Because of its importance in areas such as aerospace, implantable
 medical device, automotive, and civil infrastructure engineering,
 fatigue is the most thoroughly studied failure mode in reliability.
 Engineers routinely conduct laboratory life tests, subjecting
 material specimens to cyclic stresses that induce the initiation
 and progression of cracks or other forms of damage.  Experiments
 typically test samples of units as several fixed levels of stress
 amplitude (other experimental variables like mean stress or
 temperature are sometimes also used, but such extensions---which
 are straightforward to handle statistically---are not considered in
 this paper). Units are tested until failure (defined in some
 purposeful manner such as specimen fracture) or the end of the
 test. Unfailed units result in right-censored observations (known
 as runouts in the fatigue literature).

This section briefly reviews the statistical models used in
\citet{Meekeretal2022} and that we use for applications in this
paper.  Section~\ref{section:s.n.relationships} describes \SN{}
(stress or strain $S$ versus number of cycles to failure $N$)
relationships. Section~\ref{section:specify.fatigue.life.introduction}
reviews models for fatigue \SN{} data that use the traditional method
of specifying a fatigue-life model. The specified fatigue-life model
then induces a fatigue-strength model, as described more fully in
\citet[][Section~2]{Meekeretal2022}.
Section~\ref{section:specify.fatigue.strength.introduction} reviews
models for fatigue \SN{} data that use a new method of specifying a
fatigue-strength model which then induces a fatigue-life model, as
described more fully in \citet[][Section~3]{Meekeretal2022}. This new approach
to \SN{} model specification has important advantages in applications
where nonlinear regression is needed.

\subsection{\SN{} relationships}
\label{section:s.n.relationships}
A specified \SN{} regression relationship is the core component of any statistical model
used to describe experimental fatigue data. This relationship
describes how some quantile (usually the median) of the fatigue-life
distribution is affected by stress. Usually, the relationship is
specified in terms of a  positive monotonically decreasing
function $N=\gfun(S;\betavec)$, relating median number of cycles $N$ to
stress amplitude $S$. The function must be
monotonically decreasing because increasing stress
tends to lead to shorter lifetimes. The vector $\betavec$ contains
regression coefficients to be estimated from the available fatigue
data. Numerous nonlinear \SN{} relationships have been suggested in
the fatigue literature. Some of these are reviewed, for example, in
\citet{CastilloFernandez-Canteli2009} and \citet{Meekeretal2022}.

As explained in
\citet[][Chapters 9 and 15]{Dowling2013}, some fatigue experiments
(depending on material properties and the levels of stress used in
the experiment) are
stress-controlled while others are strain-controlled. To simplify
the presentation we will generally use the word ``stress'' to mean stress
amplitude
\textit{or} strain amplitude used as the experimental factor in a fatigue
experiment. An exception will be  in
Section~\ref{section:estimation.Inconel.data} when we present
the Inconel 718 example where strain amplitude was controlled.

The simplest \SN{} relationship
in known as the Basquin relationship
\begin{align}
\label{equation:basquin.relationship}
\log(N) &= \log[\gfun(S;\betavec)] = \beta_{0} + \beta_{1}\log(S),
\end{align}
given by \citet{Basquin1910} where log life ($\log(N$) tends to be linear in
log stress ($\log(S)$) and $\beta_{1}<0$.  In
application areas other than fatigue (e.g., accelerated life testing), this
relationship is known as the inverse power law.

More generally, the function $\gfun(S;\betavec)$, for fixed $\betavec$,
needs to be a positive monotonically decreasing because
the lifetime $N>0$ tends to decrease as stress $S>0$ increases.
In some settings (as described in
Section~\ref{section:specify.fatigue.strength.introduction}) it is
either convenient or necessary to use instead the \SN{} relationship
$S = \hfun(N;\betavec) = \gfun^{-1}(N;\betavec)$ which, for fixed
$\betavec$, is a positive monotonically decreasing function of $N$.
Especially in high-cycle fatigue tests (where units are tested at
lower levels of stress and fatigue life times tend to be long),
there can be strong curvature
(usually, but not always the curvature is concave-up over the range
of the data) in the
\SN{} relationship, requiring the use of nonlinear regression
models. The models used in
Sections~\ref{section:parameterization.initial.values.coffin.manson.model}--\ref{section:parameterization.initial.values.box.cox.loglinear.model}
illustrate such nonlinear regression relationships.

\subsection{Specifying the fatigue-life model}
\label{section:specify.fatigue.life.introduction}
When specifying a model for fatigue life $N$ at a specified level of
stress $S_{e}$, we write
\begin{align*}
  \log(N)&=\log[\gfun(S_{e};\betavec)]+\sigma_{N} \epsilon.
\end{align*}
Here $\sigma_{N} \epsilon$ is a random-error term where
$\epsilon$ has a location-scale distribution with $\mu=0$ and
$\sigma=1.$ Then for a given level of stress
amplitude $S_{e}$, the cdf of $N$ is
\begin{align}
\label{equation:fatigue.life.failure.time.model.cdf}
F_{N}(t; S_{e})&=\Pr \left(N \le t; S_{e}\right)=\Phi
         \left (\frac{\log(t)-\log[\gfun(S_{e};\betavec)]}{\sigma_{N}}   \right), \quad \quad
  t>0, \,\, S_{e}>0,
\end{align}
which is a log-location-scale distribution with scale parameter
$\gfun(S_{e};\betavec)>0$ and shape parameter $\sigma_{N}$. The
log-location-scale family of distributions includes the widely used
(for fatigue and other applications involving time-to-event data)
lognormal and Weibull distributions as special cases.

``Fatigue strength'' (which is sometimes referred
to as ``fatigue resistance'') is a random variable, denoted by $X$,
giving the level of stress that leads to a failure at a
\textit{specified number of cycles} $N_{e}$.
Using this definition of $X$ implies that the models for the
random variables $N$ and $X$ have the same random component
$\epsilon$. Although $X$ cannot be directly
observed, the distribution of $X$ can be estimated from \SN{}
data.
We note that in the fatigue literature, there are alternative
definitions of fatigue strength, but the one we use is the most
common one.

As shown in \citet[][Section~2.4.1]{Meekeretal2022}, for the
specified log-location-scale fatigue-life model
in~(\ref{equation:fatigue.life.failure.time.model.cdf}), the above
definition of $X$ implies (when $\gfun(S_{e};\betavec)$ has \textit{neither}
a vertical nor a horizontal asymptote) that the cdf of $X$ is
\begin{align}
\label{equation:general.fatigue.strength.cdf.ls} 
F_{X}(x; N_{e})&=  \Pr \left(X \le x; N_{e}\right)=   
  \Phi\left[\frac{\log(N_{e})-\log[\gfun(x;\betavec)]}{\sigma_{N}}
    \right],
\quad \quad   x>0, \,\, N_{e}>0.
\end{align}
If $N=\gfun(S;\betavec)$ has one or both coordinate asymptotes,
technical adjustments are needed to define the cdf, as described in
\citet[][Sections~2.4.2 and~2.4.3]{Meekeretal2022}.
The cdf in (\ref{equation:general.fatigue.strength.cdf.ls}) is \textit{not}
a log-location-scale distribution unless $\log[\gfun(x;\betavec)]$
is a linear function of $\log(x)$ (i.e., the Basquin relationship).

\subsection{Specifying the fatigue-strength model}
\label{section:specify.fatigue.strength.introduction}
As an alternative to the traditional method of specifying a
fatigue-life model (that will then induce a fatigue-strength model
for the random variable $X$), \citet[][Section~3.2]{Meekeretal2022} suggest
specifying a fatigue-strength model for $X$ that will then induce a
fatigue-life model for $N$. 
The biggest advantage \citep[described more
  fully in][Section~3.2.1]{Meekeretal2022} of this new approach is
that the specification of the fatigue strength model is simpler in
the frequently occurring applications where the \SN{} relationship is
nonlinear on log-log scales. In particular, the fatigue-life
distribution often has increasing spread at lower levels of
stress $S$; the fatigue-strength distributions typically have a constant
spread as a function of number of cycles $N_{e}$.

To specify a fatigue-strength distribution, let
\begin{align}
\label{equation:logx.general.strength.model}  
\log(X) = \log[\hfun(N_{e};\betavec)] +  \sigma_{X} \epsilon,
\end{align}
where $S=\hfun(N;\betavec)$ defines the \SN{} regression
relationship.  Then $\sigma_{X}
\epsilon$ is a random-error term where $\epsilon$ has a
location-scale distribution with $\mu=0$ and $\sigma=1.$ This model
implies that, for a given number of cycles $N_{e}$, $X$ has a cdf
\begin{align*}
  F_{X}(x; N_{e})&= \Pr(X \le x; N_{e}) = \Phi\left[ \frac{\log(x) -
      \log[\hfun(N_{e};\betavec)]}{\sigma_{X}}\right], \quad \quad
  x>0, \,\, N_{e}>0.
\end{align*}
Then $X$ has a log-location-scale distribution with scale parameter
$\hfun(N_{e};\betavec)>0$ and constant shape parameter~$\sigma_{X}$.

As shown in \citet[][Section~3.2.3]{Meekeretal2022}, when the
monotonically decreasing $S=\hfun(N;\betavec)$ has \textit{neither} a
vertical nor a horizontal asymptote, this model implies that
fatigue-life $N$, for a given value of stress $S_{e}$, has the cdf
\begin{align}
\label{equation:cdf.for.induced.fatigue.life} 
F_{N}(t; S_{e})&=\Pr(N \le t; S_{e})=\Phi\left[\frac{\log(S_{e})-\log[\hfun(t;\betavec)]}{\sigma_{X}}   \right], \quad \quad
  t>0, \,\, S_{e}>0.
\end{align}
If $S=\hfun(N;\betavec)$ has one or both coordinate asymptotes,
technical adjustments are needed to define the cdf, as described in
\citet[][Sections~3.2.4 and~3.2.5]{Meekeretal2022}.
The cdf in (\ref{equation:cdf.for.induced.fatigue.life}) is not a
log-location-scale distribution unless $\log[\hfun(t;\betavec)]$ is
a linear function of $\log(t)$ (i.e., the Basquin  relationship).

Because $N$ is observable (and $X$ is not), we use the
distribution of $N$ (in either
(\ref{equation:fatigue.life.failure.time.model.cdf}) or
(\ref{equation:cdf.for.induced.fatigue.life})) to define the
likelihood when fitting an \SN{} model to \SN{} data. When needed, pdfs
for either situation are obtained by using $f_{N}(t; S_{e}) =
dF_{N}(t; S_{e})/dt$.

\subsection{Scaling the data}
As suggested in Step~\ref{step:scale.data} in
Section~\ref{section:steps.ml.estimation}, it is useful and sometimes
important to scale data before doing estimation.  In our examples
involving \SN{} data, we scale both the stress/strain variable $S$
and the response variable $N$.  Dividing all values of $S$ and $N$
by the largest values of these variables (denoted by $S_{\max}$ and
$N_{\max}$, respectively) generally works well.

\subsection{Potential estimability problems}
\label{section:potential.estimability.problems}
Section~\ref{section:steps.ml.estimation}
mentioned the important
concept of SPs. Sections~\ref{section:parameterization.initial.values.coffin.manson.model}--\ref{section:parameterization.initial.values.box.cox.loglinear.model}
provide, as examples, the details the parameterizations
that we tailored for the three
five-parameter models that we have used in this paper. 
Although it is not a strict requirement for \SN{} relationships, most
\SN{} data, especially when at least some units are tested at low
levels of stress,
have concave-up curvature and increased spread at
lower levels of stress; the models used in
Sections~\ref{section:parameterization.initial.values.coffin.manson.model}--\ref{section:parameterization.initial.values.box.cox.loglinear.model}
usefully describe such data. In situations where the data plotted on
log-log scales are approximately linear (e.g.,
low-cycle fatigue where all units are exposed to relatively high levels of
stress), the Basquin relationship (\ref{equation:basquin.relationship})
may be appropriate. 

In other model/data combinations, attempting to fit a more complicated model
could lead to situations (depending on the model and definition of
the USPs) where the maximum of
the likelihood is at or near a boundary of the parameter space
or the maximum may
not be unique (due to flatness in the likelihood or profile
likelihood surface). We have
designed algorithms to detect such
situations (somewhat tailored for each
model) and provide
suitable warnings. In such situations, one should attempt to fit the indicated
limiting model (i.e., a model implied by one of the USPs approaching
$\infty$ or $-\infty$) or other special-case models (such as the
Basquin model)
and carefully check model-fitting diagnostics
(e.g., based on the residuals) to ensure that the model is
adequate. The nested model (implied by a limiting case of the
original model) can be compared with the
original model by using a likelihood-ratio test
\citep[as described in][Appendix~C.7.5]{MeekerEscobarPascual2021}.
We illustrate some of these situations in our numerical examples in
Sections~\ref{section:estimation.Inconel.data}
and~\ref{section:estimation.Polynt.data}.

\section{Application of the Procedure for the
  Coffin--Manson Model}
\label{section:parameterization.initial.values.coffin.manson.model}
\subsection{The Coffin--Manson model}
\label{subsection:steps.for.fitting.coffin.manson.model}
\begin{figure}
\begin{tabular}{cc}
(a) & (b) \\[-3.2ex]
\rsplidapdffiguresize{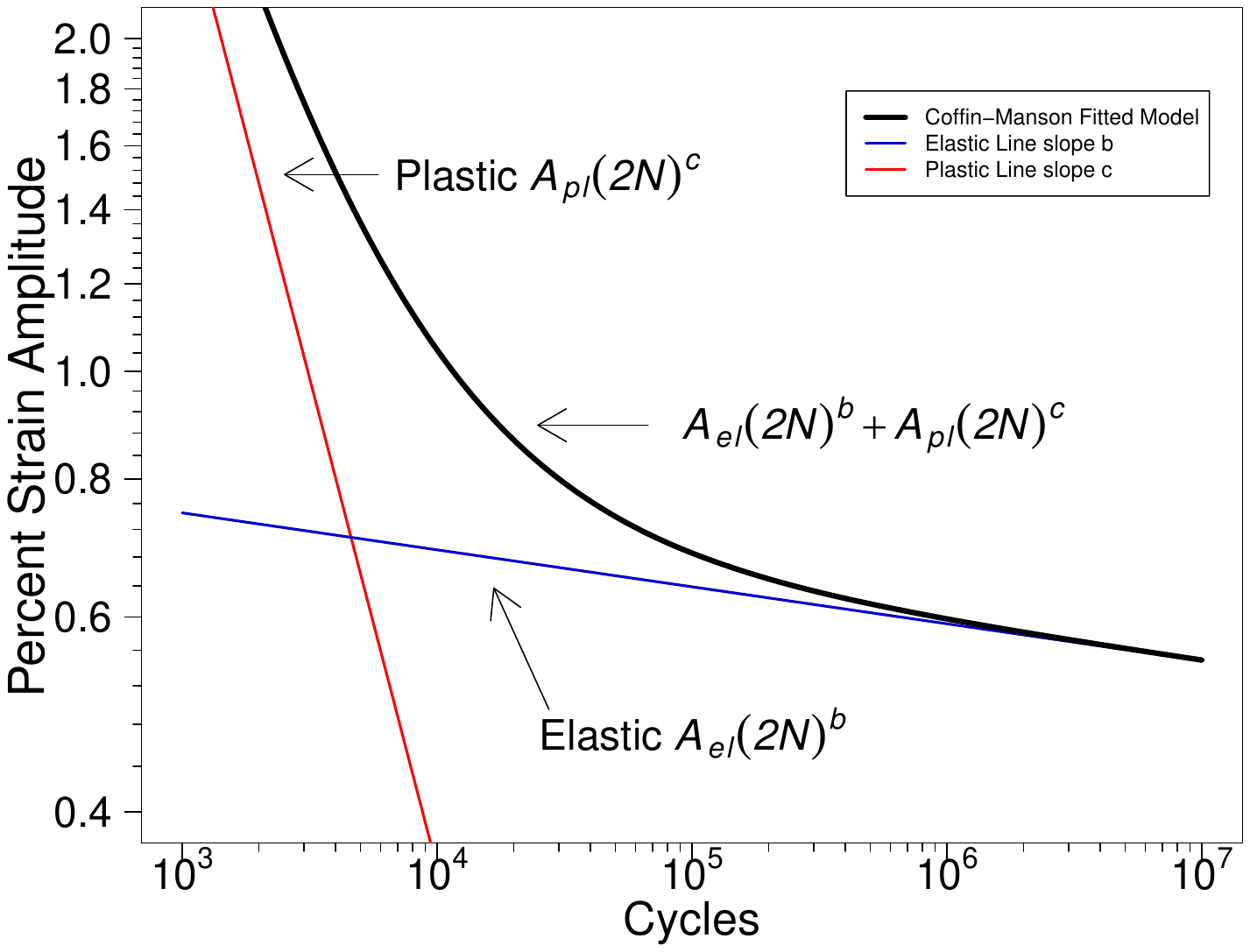}{3.25in}&
\rsplidapdffiguresize{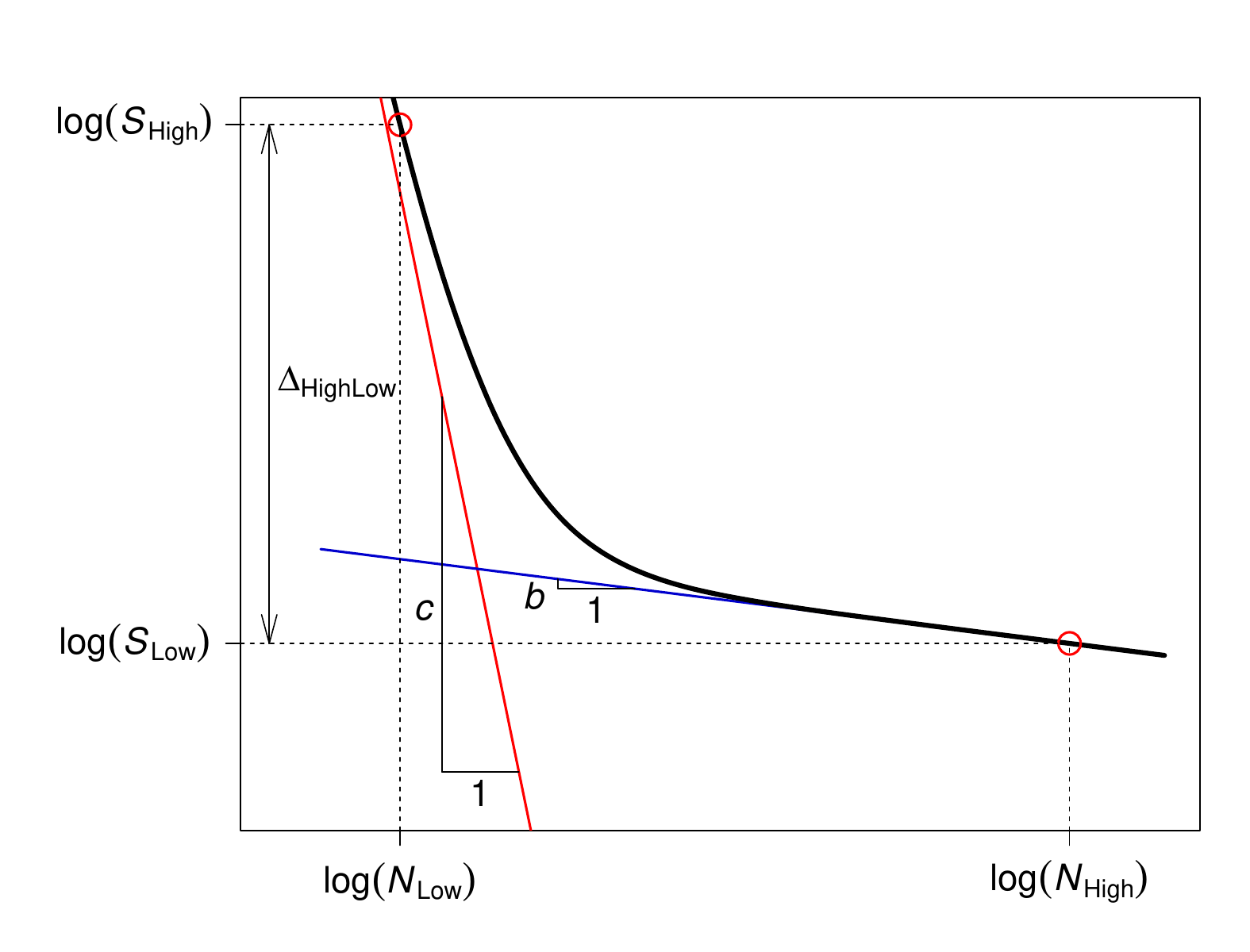}{3.25in}\\
\end{tabular}
\caption{Plot showing the Coffin--Manson \SN{} relationship~(a); Plot
  illustrating the Coffin--Manson  USPs~(b).}
\label{figure:coffin.manson.intro.plots}
\end{figure}

The Coffin--Manson relationship \citep[e.g.,][pages~724--726]{Dowling2013}
is a popular relationship for \SN{} data that exhibit,
when plotted, concave-up curvature.  The relationship, depicted
in Figure~\ref{figure:coffin.manson.intro.plots}(a),
can be interpreted as the sum
of plastic and elastic Basquin \SN{} relationships and is expressed as:
\begin{align}
\label{equation:cm.relationship}
S = h(N; \betavec)= \Ael (2N)^{b} +  \Apl (2N)^{c},
\end{align}
where the TPs are $\Ael>0$, $\Apl \geq 0$, $b \leq 0$, $c<0$, and $|c|>|b|$.
The last inequality makes the model identifiable.
It is traditional to use $2N$ in (\ref{equation:cm.relationship})
because there are two stress within each extremes cycle.

\subsection{Coffin--Manson model SPs and USPs}
\label{section:cm.stable.and.unrestricted.stable.parameterizations}
As mentioned in Section~\ref{section:steps.ml.estimation}, one way to
identify SPs is to use model characteristics that can
be easily identified from a plot of the fitted model.
The dark solid curve in
Figure~\ref{figure:coffin.manson.intro.plots}(b) is a fitted
Coffin--Manson relationship (the data have been suppressed in the
plot). The figure also illustrates several quantities that provide
the basis for our USPs.

Our Coffin--Manson SPs start with the stress levels $S_{\Lowx}$ and
$S_{\Highx}$ that correspond to the two extreme
points on the fitted \SN{}
curve in Figure~\ref{figure:coffin.manson.intro.plots}(b).
These values can be computed from the TPs by using
\begin{align}
\begin{split}
\label{equation:cm.stable.x.definition}
	S_{\Lowx} & =  \Ael(2N_{\Highx})^{b}+\Apl(2N_{\Highx})^{c}\\
	S_{\Highx} & =  \Ael(2N_{\Lowx})^{b}+\Apl(2N_{\Lowx})^{c},
\end{split}
\end{align}
where $N_{\Highx}$ is the largest value of $N$ in the data that is a
failure and
$N_{\Lowx}$ is the smallest value of $N$ in the data.
We also use the elastic-line slope $b$ and the plastic-line slope
$c$ as SPs.
The vertical spread in the data points about the fitted model
provides an indication of the value of $\sigma_{X}$ and thus $\sigma_{X}$
is an SP (even though fatigue strength $X$
is not directly observable and applied stress $S$ is an experimental
factor, the vertical spread in the \SN{} data provides an
indication of the spread in distribution of fatigue strength).

The SPs have the following constraints and order restrictions:
\begin{itemize}
\item
$S_{\Highx} > S_{\Lowx} > 0$,
\item
$c < \limitSlope < b < 0$, and 
\item
$\sigma_{X} > 0 $,
\end{itemize}
where
\begin{align}
\begin{split}
  \label{equation:cm.slope.limit}
\limitSlope & = \frac{\log(S_{\Lowx})-\log(S_{\Highx})}{\log(N_{\Highx})-\log(N_{\Lowx})}
\end{split}
\end{align}
is the slope of the line that connects the two open-circle points
in Figure~\ref{figure:coffin.manson.intro.plots}(b).
Our corresponding USPs are
\begin{align}
\begin{split}
  \label{equation:cm.unrestricted.stable.definition}
& \logSLow  = \log(S_{\Lowx}) \\
& \logDeltaHighLow = \log[\log(S_{\Highx}) - \log(S_{\Lowx})] \\
  &  \qlogisp = \qlogis(b/\limitSlope)\\
  & \logDeltaSlopes  =  \log\left(\limitSlope-c\right), \textrm{ and}\\
& \logSigmaX=\log(\sigma_{X}),
\end{split}
\end{align}
where $\qlogis$ is the standard logistic distribution quantile
function (also known as the logit transformation).

The order restrictions $c < \limitSlope < b$ ensure that the resulting
model is concave-up (and, correspondingly, that $\Ael$ and $\Apl$
will be positive).  Analytically, this can be seen as follows.  The
solutions of (\ref{equation:cm.stable.x.definition}) for $\Ael$ and
$\Apl$ are:
\begin{align}
\begin{split}
\label{equation:cm.eqn.solutions}
	\Ael & = \dfrac{S_{\Lowx}/(2N_{\Highx})^{c}  -
          S_{\Highx}/(2N_{\Lowx})^{c}}
         {(2N_{\Highx})^{b-c} 
          - (2N_{\Lowx})^{b-c}}\\
	\Apl & = \dfrac{S_{\Lowx}/(2N_{\Highx})^{b}  -
          S_{\Highx}/(2N_{\Lowx})^{b}}
         {(2N_{\Highx})^{c-b}
          - (2N_{\Lowx})^{c-b}}.
\end{split}
\end{align}
In (\ref{equation:cm.eqn.solutions}), because $N_{\Highx}>N_{\Lowx}$ and $c<b$, 
the denominator of the solution for $\Ael$ is positive
and the denominator of the solution for $\Apl$ is negative.
Because both $\Ael$ and $\Apl$ must be positive,
the numerator of the solution for $\Ael$ must be positive,
and the numerator of the solution for $\Apl$ must be negative,
as expressed in the following inequalities:
\begin{align}
\begin{split}
\label{equation:cm.slope.inequalities}
	S_{\Lowx}/(2N_{\Highx})^{c}  - S_{\Highx}/(2N_{\Lowx})^{c} &> 0\\
	S_{\Lowx}/(2N_{\Highx})^{b}  - S_{\Highx}/(2N_{\Lowx})^{b} &< 0.
\end{split}
\end{align}
Solving (\ref{equation:cm.slope.inequalities}) for $b$ and $c$
leads to the order restrictions
$c < \limitSlope < b$ and the expression for $\limitSlope$ in~(\ref{equation:cm.slope.limit}).

\subsection{Initial values for the Coffin--Manson USPs}
\label{section:cm.initial.values.traditional.parameters}
Our algorithm determines initial values for the SPs first, then
translates those initial values into initial values for the USPs.
An optimization routine then finds the values of the USPs
that maximize the likelihood function.

The initial value for $S_{\Lowx}$ is the smallest stress value among all
failure observations (i.e., omitting the right-censored observations).
The initial value for $S_{\Highx}$ is the largest stress
value among all failure observations.
The initial values for $b$ and $c$ are ordinary least
squares (OLS) slope estimates of two
simple linear regressions using two data subsets.
The two data subsets are obtained by dividing the data
into two groups (below and above the median lifetime)
and treating right-censored observations as failures.
For each data subset, a simple linear regression is fit
using OLS
using the logarithm of stress as
the response variable and the logarithm of lifetime
as the explanatory variable.
The slope estimate from the subset with smaller lifetimes is the
initial value for  $c$, the elastic-line slope.
The other slope estimate is for $b$, the plastic-line slope.

An initial value for $\sigma_{X}$ is obtained by using the
residual standard deviation from an OLS fit, again using the
logarithm of stress as the response variable and the logarithm of
lifetime as the explanatory variable. This results in an estimate
that is biased high, which we have found to be a robust initial
value for $\sigma_{X}$. Then one computes the initial values of the USPs
from (\ref{equation:cm.slope.limit})
and (\ref{equation:cm.unrestricted.stable.definition}).

\subsection{Computing the Coffin--Manson TPs from
  the USPs}
\label{section:cm.traditional.parameters.from.stable}
The equations for the likelihood itself will usually be programmed
in terms of the TPs. Before computing the
likelihood, one then needs to compute the TPs from
the USPs, usually by calling a function to do this at the very
beginning of the function to compute the likelihood.
Also, when estimation is complete there is need to compute the ML
estimates of the TPs.
The Coffin--Manson TPs as functions of our
Coffin--Manson USPs are:
\begin{align}
\begin{split}
  \label{equation:cm.stable.to.traditional}
  b & =  \limitSlope \cdot \plogis(\qlogisp)\\[0.5ex]
  c & =  \limitSlope - \exp(\logDeltaSlopes)\\[0.5ex]
  \Apl & =  \dfrac{S_{\Lowx}/(2N_{\Highx})^{b}  -
          S_{\Highx}/(2N_{\Lowx})^{b}}
         {(2N_{\Highx})^{c-b}
          - (2N_{\Lowx})^{c-b}}\\[0.5ex]
  \Ael & = \dfrac{S_{\Highx}}{(2N_{\Lowx})^{b}} -
           (2N_{\Lowx})^{c-b}\Apl\\[0.5ex]
  \sigma_{X} &= \exp(\logSigmaX),
\end{split}
\end{align}
where $\plogis$ is the standard logistic distribution cdf
(also known as the inverse logit transformation),
\begin{align*}
  S_{\Lowx} &= \exp(\logSLow) \\[0.5ex]
  S_{\Highx} &= \exp(\logSLow + \exp[\logDeltaHighLow]),
\end{align*}
and $\limitSlope$ is computed using (\ref{equation:cm.slope.limit}).

\subsection{ML estimates of the traditional parameters
  based on the original unscaled data}
As described in Step~\ref{step:scale.data} of
Section~\ref{section:steps.ml.estimation}, to ensure
numerical stability, we use the scaled data to calculate the
likelihood function as a function of the USPs.
Thus the initial results of ML estimation provide ML estimates of the USPs
based on the scaled data. To recover the ML estimates
for TPs based on the scaled data,
one uses (\ref{equation:cm.stable.to.traditional}). 

To report final results the ML estimates
of the TPs for the original \textit{unscaled}
data (which we call TPNSs) can be computed as follows.
Denote the ML estimates for the TPs from the scaled data
by $\widetilde{A}_{pl}$, $\widetilde{b}$, $\widetilde{A}_{el}$,
and $\widetilde{c}$,
where the scaling values are $S_{\max}$ and $N_{\max}$ respectively for
stress and number of cycles. Then the fitted Coffin--Manson model
for the scaled data is:
\begin{align*}
\widetilde{S} = \widetilde{A}_\textrm{el}(2\widetilde{N})^{\widetilde{b}}+\widetilde{A}_\textrm{pl}(2\widetilde{N})^{\widetilde{c}},
\end{align*}
where $\widetilde{S}$ and $\widetilde{N}$ are scaled stress and
lifetime values.
In terms of the original data and scaling factor, the relationship is:
\begin{align*}
	S = S_{\max}\Atildeel N_{\max}^{-\widetilde{b}}(2N)^{\widetilde{b}}
          + S_{\max}\Atildepl N_{\max}^{-\widetilde{c}}(2N)^{\widetilde{c}}.
\end{align*}
Therefore, the ML estimates for the TPNSs are:
\begin{align*}
	\Ahatel & = S_{\max} \Atildeel N_{\max}^{-\widetilde{b}}\\
	\Ahatpl & = S_{\max} \Atildepl N_{\max}^{-\widetilde{c}}\\
	\bhat & = \widetilde{b}\\
	\chat & = \widetilde{c}.
\end{align*}

\subsection{Coffin--Manson limiting cases and competing models}
\label{section:coffin.manson.limiting.case}
The Coffin--Manson model has two limiting cases that are plausible \SN{}-curve  models. 
One is the Coffin--Manson Zero-Elastic-Slope model (where the elastic
line becomes a horizontal asymptote), and the other is the Basquin model.
In terms of TPs, when the parameter $b$ in (\ref{equation:cm.relationship}) approaches zero,
the Coffin--Manson model approaches the Coffin--Manson Zero-Elastic-Slope model.
The limiting model has the following \SN{} relationship:
\begin{align}
\label{equation:cm.zero.elastic.slope.relationship}
S = h(N; \betavec)= \Ael +  \Apl (2N)^{c}.
\end{align}
The second limiting model is the Basquin model.
When either $\Ael \rightarrow 0$ or $\Apl \rightarrow 0$,
(\ref{equation:cm.relationship}) approaches the following \SN{} relationship:
\begin{align}
\label{equation:basquin.sn.relationship}
S = h(N; \betavec)= \A (2N)^{\limitSlope}.
\end{align}
after reparameterization,  (\ref{equation:basquin.relationship})
and (\ref{equation:basquin.sn.relationship}) are equivalent.

In terms of the USPs, when parameter $\qlogis(p) \rightarrow -\infty$
the Coffin--Manson model approaches
the Coffin--Manson Zero-Elastic-Slope model~(\ref{equation:cm.zero.elastic.slope.relationship})
(because $b \rightarrow  0$).
When $\log(\delta) \rightarrow 0$ or $\qlogis(p) \rightarrow \infty$
or both,
the Coffin--Manson model approaches the Basquin model.
This can be seen by letting $\log(\delta) \rightarrow 0$ which
implies that $c \rightarrow \limitSlope$
and letting $\qlogis(p) \rightarrow \infty$ which implies that
$b \rightarrow  \limitSlope$.
Therefore, $\Apl \rightarrow  0$ or $\Ael \rightarrow  0$, according to
(\ref{equation:cm.eqn.solutions}) and (\ref{equation:cm.slope.inequalities}) in
Section~\ref{section:cm.stable.and.unrestricted.stable.parameterizations}.

\section{Application of the Procedure for the Nishijima Model}
\label{section:parameterization.initial.values.nishijima.hyperbolic.model}
\subsection{The Nishijima model}

\begin{figure}
\begin{tabular}{cc}
(a) & (b) \\[-3.2ex]
\rsplidapdffiguresize{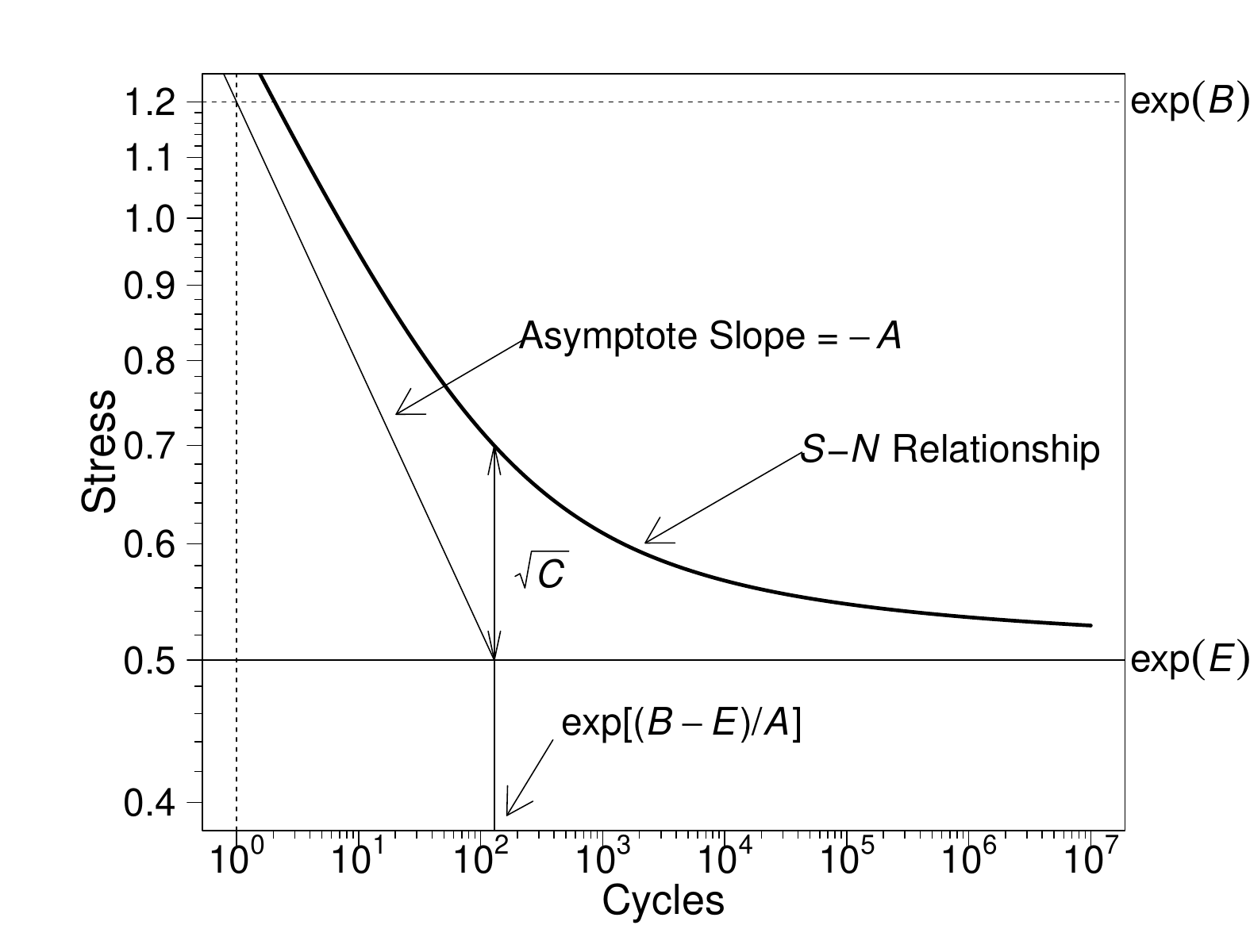}{3.25in}&
\rsplidapdffiguresize{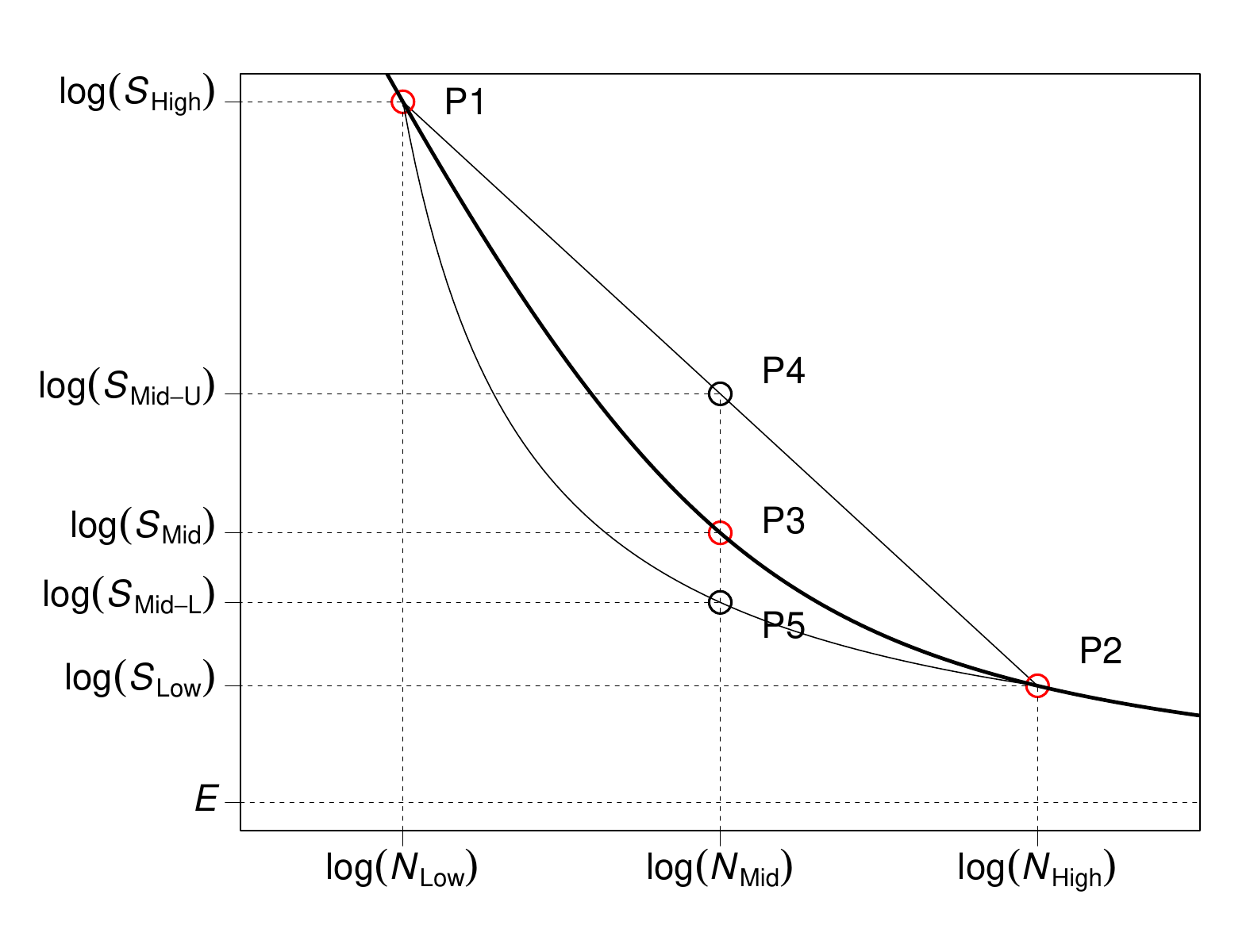}{3.25in}\\
\end{tabular}
\caption{Plot showing the Nishijima \SN{} relationship~(a); Plot
  illustrating the Nishijima USPs~(b).}
\label{figure:nishijima.intro.plots}
\end{figure}

The  Nishijima \SN{} relationship is described and used in
\citet{Falk2019} and \citet{Meekeretal2022}. 
Although it is less well known than the Coffin--Manson model,
the Nishijima model provides a good description of many \SN{} data sets.
The Nishijima relationship is the hyperbola:
\begin{align}
\label{equation:nishijima.hyperbolic.relationship}
[\log(S)-E][\log(S)+A\log(N)-B]&= C,
\end{align}
or equivalently
\begin{align}
\label{equation:nishijima.hyperbolic.relationship.hfunction.for.numerical}
S&= h(N; \betavec) = \exp\left(\frac{
	-A \log(N)+B+E + \sqrt{
		\left[A \log(N)-(B-E)\right]^{2} +4C}}{2}\right),
\end{align}
depicted
in Figure~\ref{figure:nishijima.intro.plots}(a),
where $A$, $B$, $C$, and $E$  (indicated
on the plot) are the regression model TPs. The TPs have the constraints:
\begin{displaymath}
\begin{array}{cccc}
	& A & > & 0 ,\\
	& C & > & 0 .
\end{array}
\end{displaymath}
These constraints ensure that 
(\ref{equation:nishijima.hyperbolic.relationship.hfunction.for.numerical})
is concave-up.
The parameter $A$ is the negative of slope of the oblique asymptote,
$B$ is the $\log(N)=0$ intercept of the same asymptote, $E$ is the
log of the horizontal asymptote, and $\sqrt{C}$ is the vertical
distance between the \SN{} curve and the point where the two
asymptotes intersect.

\subsection{Nishijima model SPs and USPs}
Similar to identifying model characteristics for the Coffin--Manson model,
the stable parameterization for the Nishijima model can be observed from
the plot of a fitted model, illustrated by
Figure~\ref{figure:nishijima.intro.plots}(b).
The SPs of the Nishijima model consist of $S_{\Highx}$,
$S_{\Midx}$, $S_{\Lowx}$, $E$, and $\sigma_X$. The first three SPs,
which replace the traditional $A$, $B$, and $C$, are stress levels that
correspond to the three open-circle points (P1, P2, and P3)
on the thick solid Nishijima \SN{} curve in
Figure~\ref{figure:nishijima.intro.plots}(b).
These values can be computed from the traditional parameters, according to
(\ref{equation:nishijima.hyperbolic.relationship.hfunction.for.numerical}),
by using the following three points on the Nishijima relationship:
\begin{align}
\begin{split}
\label{equation:nishijima.stable.x.definition}
  S_{\Lowx} & =  h(N_{\Highx};\beta) \\
  S_{\Midx} & = h(N_{\Midx};\beta) \\
  S_{\Highx} & =  h(N_{\Lowx};\beta),
\end{split}
\end{align}
where $N_{\Lowx}$ is the smallest value of $N$ in the data,
$N_{\Highx}$ is the largest value of $N$ in the data that is a
failure, and $N_{\Midx} =
\exp[(\log(N_{\Lowx}) + \log(N_{\Highx}))/2]$ is the midpoint on
the logarithmic scale.

The SPs have the following constraints and order restrictions:
\begin{itemize}
\item $0 < \exp(E) < S_{\Lowx} < S_{\MidL} < S_{\Midx} < S_{\MidU} < S_{\Highx}$, and
\item $\sigma_X>0$,
\end{itemize}
where
\begin{align}
\begin{split}
\label{equation:nishijima.stable.x.MidU.MidL}
  \log(S_{\MidU}) & =  \frac{1}{2} \,[\log(S_{\Highx})+\log(S_{\Lowx})] \\
  \log(S_{\MidL}) & =  2 \, \frac{[\log(S_{\Lowx})-E][\log(S_{\Highx})-E]}{\log(S_{\Lowx})+\log(S_{\Highx})-2E}+E.
\end{split}
\end{align}
The open-circles P3, P4, and P5
in Figure~\ref{figure:nishijima.intro.plots}(b)
have their horizontal position at $\log(N_{\Midx})$.
The stress values $S_{\MidU}$ and $S_{\MidL}$
give the vertical positions of the the open-circles P4 and P5.
The open-circle P4 is at the middle of the line segment connecting
P1 and P2, corresponding to the limiting linear \SN{} relationship.
The open-circle P5 is on a curve that corresponds to the limiting
rectangular hyperbola containing points P1 and P2
and having $E$ as its horizontal asymptote. 
The order restriction $S_{\Midx} < S_{\MidU}$ ensures that the Nishijima
\SN{} curve is concave-up and can approach the
linear \SN{} relationship between P1 and P2.
The order restriction $S_{\MidL} < S_{\Midx}$
ensures that the Nishijima \SN{} curve is a
hyperbola that can approach the
limiting Rectangular Hyperbola \SN{} relationship.
Section~\ref{section:nishijima.limiting.case} describes both limits.

To derive the order restrictions $S_{\MidL} < S_{\Midx} < S_{\MidU}$,
start with the following equations:
\begin{align}
\begin{split}
\label{equation:nishijima.stable.x.definition.linear.system}
  -A\log(N_{\Highx}) + B + \frac{C}{\log(S_{\Lowx})-E} & = \log(S_{\Lowx}) \\
  -A\log(N_{\Midx}) + B + \frac{C}{\log(S_{\Midx})-E} & = \log(S_{\Midx}) \\
  -A\log(N_{\Lowx}) + B + \frac{C}{\log(S_{\Highx})-E} & = \log(S_{\Highx}).
\end{split}
\end{align}
These equations are are obtained from
(\ref{equation:nishijima.stable.x.definition}).
For example, to convert the first equation in
(\ref{equation:nishijima.stable.x.definition})
to the first equation in
(\ref{equation:nishijima.stable.x.definition.linear.system}),
use (\ref{equation:nishijima.hyperbolic.relationship}) to
rewrite the first equation in (\ref{equation:nishijima.stable.x.definition})
as follows:
\begin{align}
\label{equation:nishijima.hyperbolic.relationship.SLow}
[\log(S_{\Lowx})-E][\log(S_{\Lowx})+A\log(N_{\Highx})-B]&= C.
\end{align}
Then divide both sides by $[\log(S_{\Lowx})-E]$ and move the
terms involving $A$, $B$, $C$, and $E$ to the same side of the equation.
Now add the first and the third equations in
(\ref{equation:nishijima.stable.x.definition.linear.system}) together
and subtract two times the second equation. The resulting
equation simplifies because
the terms involving $A$ and $B$
either cancel out or because $\log(N_{\Midx})$
is the midpoint of the line between $\log(N_{\Lowx})$ and $\log(N_{\Highx})$.
The final result is:
\begin{align}
\label{equation:nishijima.hyperbolic.relationship.SPs.relation}
  \log(S_{\Midx}) - \frac{1}{2}\left[\log(S_{\Lowx})+\log(S_{\Highx})\right]
  &=C\left(\frac{1}{\log(S_{\Midx})-E}-\frac{1}{2}\, \frac{\log(S_{\Lowx})+\log(S_{\Highx})-2E}{[\log(S_{\Highx})-E][\log(S_{\Lowx})-E]}\right).
\end{align}

The left hand side of
(\ref{equation:nishijima.hyperbolic.relationship.SPs.relation})
is negative because of $S_{\Midx} < S_{\MidU}$. Then because $C>0$,
\begin{align*}
  \left(\frac{1}{\log(S_{\Midx})-E}-\frac{1}{2}\, \frac{\log(S_{\Lowx})+\log(S_{\Highx})-2E}{[\log(S_{\Highx})-E][\log(S_{\Lowx})-E]}\right) & < 0.
\end{align*}
After rearranging terms, one can see that 
$S_{\MidL}<S_{\Midx}$, where $S_{\MidL}$ is  given by the second
equation in (\ref{equation:nishijima.stable.x.MidU.MidL}).

Because the SPs have order restrictions and
constraints, the corresponding Nishijima USPs are
\begin{align}
\begin{split}
  \label{equation:nishijima.unrestricted.stable.definition}
& \logSLow  = \log(S_{\Lowx}) \\
& \logDeltaHighLow = \log[\log(S_{\Highx}) - \log(S_{\Lowx})] \\
& \qlogisp = \qlogis(p) \\
& \logDeltaSLowE = \log[\log(S_{\Lowx}) - E] \\
& \logSigmaX=\log(\sigma_{X}),
\end{split}
\end{align}
where $\qlogis$ is the standard logistic distribution quantile
function (also known as the logit transformation),
\begin{align*}
p &=
    \frac{\log(S_{\MidU})  -\log(S_{\Midx})}{\log(S_{\MidU})-\log(S_{\MidL})},
\end{align*}
and $\log(S_{\MidL})$ and $\log(S_{\MidU})$ are computed from
the other USPs
using (\ref{equation:nishijima.stable.x.MidU.MidL}).

\subsection{Initial values for the Nishijima USPs}
The initial values for
$S_{\Lowx}$, $S_{\Highx}$, and $\sigma_X$ are obtained in the same
way as the Coffin--Manson for the SPs with the same names.
For the other two SPs, we use:
\begin{align*}
  E & = \log(S_{\Lowx}) - 0.1[\log(S_{\Highx})-\log(S_{\Lowx})]\\
  S_{\Midx} & = \exp\left(\frac{1}{2}[\log(S_{\MidL})+\log(S_{\MidU})]\right),
\end{align*}
where the formulas for $\log(S_{\MidL})$ and $\log(S_{\MidU})$
are given in (\ref{equation:nishijima.stable.x.MidU.MidL}).
Then the initial values of USPs are calculated using
(\ref{equation:nishijima.unrestricted.stable.definition}).

\subsection{Computing the Nishijima TPs from the USPs}
\label{subsection:nishijima.usps.to.traditional}
Similar to the Coffin--Manson model described in
Section~\ref{section:cm.traditional.parameters.from.stable},
the Nishijima TPs can be computed from the Nishijima
USPs. First, compute Nishijima SPs from USPs by inverting
the operations in (\ref{equation:nishijima.unrestricted.stable.definition}).
In particular, use:
\begin{align}
\begin{split}
  \label{equation:nishijima.USPs.to.SPs}
  S_{\Lowx} &= \exp(\logSLow) \\
  E &= \logSLow - \exp(\logDeltaSLowE) \\
  S_{\Highx} &= \exp[\logSLow+ \exp(\logDeltaHighLow)] \\
  S_{\Midx} &= \exp(\log(S_{\MidU}) - \plogis(\qlogisp)\left[\log(S_{\MidU})-\log(S_{\MidL})\right]) \\
  \sigma_{X} &= \exp(\logSigmaX),
\end{split}
\end{align}
 where $\plogis$ is the standard logistic distribution cdf
(also known as the inverse logit transformation)
and $\log(S_{\MidU})$ and $\log(S_{\MidL})$ are computed from
$\log(S_{\Lowx})$, $\log(S_{\Highx})$, and $E$ using
(\ref{equation:nishijima.stable.x.MidU.MidL}).
Then these SPs are substituted into
(\ref{equation:nishijima.stable.x.definition.linear.system});
solving  that system of linear equation
provides the TPs $A$, $B$, and $C$.

\subsection{ML estimates for the TPNSs based on the original unscaled data}
The ML estimates for the TPs based on the \textit{scaled} data
are functions of the ML estimates for USPs, according to
Section~\ref{subsection:nishijima.usps.to.traditional}.
Denote these ML estimates based on the scaled data by $\widetilde{A}$,
$\widetilde{B}$, $\widetilde{C}$, and $\widetilde{E}$,
where the scaling values are
$S_{\max}$ and $N_{\max}$ for stress and number of cycles, respectively.
Then the fitted Nishijima model for the scaled data is:
\begin{align*}
\widetilde{S}= \exp\left(\frac{
	-\Atilde \log(\widetilde{N})+\Btilde+\Etilde + \sqrt{
		\left[\Atilde \log(\widetilde{N})-(\Btilde-\Etilde)\right]^{2} +4\Ctilde}}{2}\right),
\end{align*}
where $\widetilde{S}$ and $\widetilde{N}$ are scaled stress and
lifetime values.
In terms of the original (\textit{unscaled})
data and scaling factor, the relationship is:
\begin{align*}
  S & = \exp\left(\log(S_{\max})+\frac{
	-\Atilde \log\left(\frac{N}{N_{\max}}\right)+\Btilde+\Etilde + \sqrt{
      \left[\Atilde \log\left(\frac{N}{N_{\max}}\right)-
      (\Btilde-\Etilde)\right]^{2} +4\Ctilde}}{2}\right) \\
  &=\exp\left(\frac{
	-\Ahat \log(N)+\Bhat+\Ehat + \sqrt{
            \left[\Ahat \log(N)-(\Bhat-\Ehat)\right]^{2} +4\Chat}}{2}\right), \\
\end{align*}
where
\begin{align}
\begin{split}
\label{equation:nishijima.recover.original.parameters}
	\Ahat & = \Atilde \\
	\Bhat & = \Btilde + \log(S_{\max}) + \Atilde \log(N_{\max})\\
	\Chat & = \Ctilde \\
	\Ehat & = \Etilde + \log(S_{\max})\\
\end{split}
\end{align}
are the ML estimates for the TPNSs of the Nishijima model
based on the original unscaled data.

\subsection{Nishijima limiting cases and competing models}
\label{section:nishijima.limiting.case}
We have shown that the Nishijima SPs have order restrictions.
The order restrictions $S_{\MidL} < S_{\Midx} < S_{\MidU}$ are related
to two limiting cases of the Nishijima relationship. One is a Piecewise-Linear
relationship and the other is the Rectangular Hyperbola relationship.

When $S_{\Midx}$ approaches $S_{\MidU}$ from below, the left hand side of
(\ref{equation:nishijima.hyperbolic.relationship.SPs.relation})
approaches zero. Then $C$ must approach zero to ensure the equality
in (\ref{equation:nishijima.hyperbolic.relationship.SPs.relation})
holds for any values of $\log(S_{\Lowx})$ and $\log(S_{\Highx})$.
If $C \rightarrow 0$, then $\log(S)+A\log(N)-B \rightarrow 0$, according to
(\ref{equation:nishijima.hyperbolic.relationship}), for any value of
$\log(S)$.  This implies, conditioning on $\log(S)>E$, that
$\log(S)$ is approximately linear in $\log(N)$.  Thus the \SN{}
relationship of this limiting case is approximately a straight line
between P1 and P2. The approximate linear relationship continues
until it reaches the horizontal asymptote at $E$, after which it
takes a sharp turn and is again approximately a straight line but
close to the horizontal asymptote $E$.
When the value of $E$ is far from the data, the
limiting model is a straight line and suggests the Basquin
relationship as a plausible alternative.

When $S_{\Midx}$ approaches $S_{\MidL}$ from above,
the Nishijima \SN{} curve approaches a Rectangular Hyperbola \SN{} curve.
The Rectangular Hyperbola model has the following relationship:
\begin{align}
\label{equation:rectangular.hyperbolic.relationship}
[\log(S)-E^*][\log(N)-B^*]&= C^*,
\end{align}
where we use asterisk superscripts to differentiate
the limiting values of the parameters from
the similarly named parameters in the Nishijima model.
The following steps prove the result.

First, when $S_{\Midx}$ approaches $S_{\MidL}$
from above, $C$ must approach $\infty$,
according to (\ref{equation:nishijima.hyperbolic.relationship.SPs.relation}),
because its left hand side approaches a constant, while the
expression in the outer parentheses on the right-hand side of
(\ref{equation:nishijima.hyperbolic.relationship.SPs.relation})
approaches zero.

Then rewrite (\ref{equation:nishijima.hyperbolic.relationship}),
by dividing both sides by $A$ giving:
\begin{align}
\label{equation:nishijima.rectangular.hyperbolic.TP}
  [\log(S)-E] \left[\frac{1}{A} + \log(N) - \frac{B}{A}\right] = \frac{C}{A}.
\end{align}
Next, we will show that as $S_{\Midx}$ approaches $S_{\MidL}$ from above,
$1/A \rightarrow 0$, $B/A \rightarrow B^*$, and $C/A \rightarrow C^*$.

Using the first and the third equations in
(\ref{equation:nishijima.stable.x.definition.linear.system})
gives a system of linear equations in $A$ and $B$ as follows:
\begin{align}
\begin{split}
\label{equation:nishijima.stable.x.definition.linear.system.13}
  -A\log(N_{\Highx}) + B + \frac{C}{\log(S_{\Lowx})-E} & = \log(S_{\Lowx}) \\
  -A\log(N_{\Lowx}) + B + \frac{C}{\log(S_{\Highx})-E} & = \log(S_{\Highx}).
\end{split}
\end{align}
Solving for $A$ and $B$ for given values of the other quantities
gives:
\begin{align}
\begin{split}
\label{equation:nishijima.SP.ABLimits}
  A & = \textrm{constant}_A + C \, \frac{\log(S_{\Highx})-\log(S_{\Lowx})}{[\log(N_{\Highx})-\log(N_{\Lowx})][\log(S_{\Highx})-E][\log(S_{\Lowx})-E]}\\
  B & = \textrm{constant}_B + C \, \frac{\log(N_{\Lowx})[\log(S_{\Highx})-E]-\log(N_{\Highx})[\log(S_{\Lowx})-E]}{[\log(N_{\Highx})-\log(N_{\Lowx})][\log(S_{\Highx})-E][\log(S_{\Lowx})-E]},
\end{split}
\end{align}
where $\textrm{constant}_A$ and $\textrm{constant}_B$ are constants that
do not involve $A$, $B$, or $C$. When $S_{\Midx}$ approaches $S_{\MidL}$ from above,
$A$ must approach $\infty$, and $B$ must approach either $\infty$ or
$-\infty$, because $C$ approaches $\infty$.
By substituting $A$ and $B$ in (\ref{equation:nishijima.rectangular.hyperbolic.TP})
by their expressions in terms of other quantities as in
(\ref{equation:nishijima.SP.ABLimits}), gives the
following limits when $S_{\Midx}$ approaches $S_{\MidL}$:
\begin{align}
\begin{split}
\label{equation:nishijima.rectangular.hyperbola.coeff.limits}
  \frac{1}{A} & \to 0 \\[0.7ex]
  \frac{B}{A} & \to B^* = \frac{\log(N_{\Lowx})[\log(S_{\Highx})-E] - \log(N_{\Highx})[\log(S_{\Lowx})-E]}{\log(S_{\Highx})-\log(S_{\Lowx})} \\[0.5ex]
  \frac{C}{A} & \to C^* = [\log(S_{\Highx})-E][\log(N_{\Lowx})-B^*].
\end{split}
\end{align}
Therefore,  when $S_{\Midx}$ approaches $S_{\MidL}$ from above,
the oblique asymptote approaches being vertical, and
the Nishijima \SN{} relationship (\ref{equation:nishijima.rectangular.hyperbolic.TP})
approaches the Rectangular Hyperbola \SN{} relationship
(\ref{equation:rectangular.hyperbolic.relationship}) with $E=E^*$.

\section{Application of the Procedure for the
  Box--Cox/Loglinear-\texorpdfstring{$\sigma_{N}$}{Lg} Model}
\label{section:parameterization.initial.values.box.cox.loglinear.model}

\subsection{The Box--Cox/Loglinear-\texorpdfstring{$\sigma_{N}$}{Lg} model}

In the pioneering work on the statistical modeling of high-cycle
fatigue data, \citet{Nelson1984} used a quadratic model to describe
the curvature in the \SN{} relationship and a log-linear model
component to
describe the increased spread in fatigue-life at lower levels of
stress. \citet[][Chapter 17]{MeekerEscobarPascual2021} and
\citet{MeekerEscobarZayac2003}
used the Box--Cox relationship to describe
the curvature in the \SN{} relationship.
Because it is monotone decreasing, the Box--Cox model provides a better
alternative to the quadratic description of the
curvature and can be thought of as an extension of the Basquin model
where the log transformation is replaced by a Box--Cox
power transformation on stress. When the power parameter is negative,
the curvature in the \SN{} relationship is concave-up,
as is commonly seen in fatigue-life data.

In this section \citep[following][Example 2.2]{Meekeretal2022}, we
use the traditional method of \textit{specifying a fatigue life model} using
a Box--Cox \SN{} relationship with a log-linear model component for the
fatigue-life distribution shape parameter. This fatigue-life model then induces
the fatigue-strength model. We note, however, that one could also
use the Box--Cox \SN{} relationship to specify a fatigue-strength
model which would induce a fatigue-life model. In data sets we have
seen, the log-linear model component to describe the increased
spread in fatigue-life at lower levels of stress would not be
needed if this latter approach is used.
This is because\citep[as described in][Section~5.1]{Meekeretal2022},
with concave-up curvature in the \SN{} relationship, increased
spread in the fatigue life model would be a feature of the induced
model.

The following is the \SN{} median curve defined for the traditional
Box--Cox/Loglinear-$\sigma_{N}$ parameterization:
\begin{align*}
  \log(N)&= \beta_{0}+ \beta_{1} \nu(S;\lambda)
           + \exp\left[\beta^{[\sigma]}_{0} + \beta^{[\sigma]}_{1} \log(S)\right]\Phi^{-1}(0.50),
\end{align*}
where
\begin{align*}
\nu(S;\lambda) &=
\begin{cases}
\dfrac{S^{\lambda} -1}{\lambda} & \text{if $\lambda \ne  0$}\\[2ex]
\log(S) & \text{if $\lambda=0$},
\end{cases}
\end{align*}
is the Box--Cox power transformation, and $\Phi^{-1}(0.50)$ is the
median of the corresponding standard location-scale distribution.

\subsection{Box--Cox/Loglinear-\texorpdfstring{$\sigma_{N}$}{Lg} model SPs and USPs}
We suggest a parameterization that includes the four SPs that are
one-to-one functions of the traditional regression coefficients
($\beta_{0}$, $\beta_{1}$, $\beta^{[\sigma]}_{0}$,
$\beta^{[\sigma]}_{1}$).  First, define $S_{\Highx}$ to be the
highest level of stress in the data set and $S_{\Lowx}$ to be the
lowest level of stress in the data set where failures occurred (note
that, unlike the usage in
Sections~\ref{section:parameterization.initial.values.coffin.manson.model}
and~\ref{section:parameterization.initial.values.nishijima.hyperbolic.model},
these are not SPs in this section).  In particular,
\begin{align}
\begin{split}
\label{equation:s.low.definition}
	\sigma_{\Lowx} & =  \exp[\beta^{[\sigma]}_{0}+\beta^{[\sigma]}_{1}\log(S_{\Highx})]\\
	\sigma_{\Highx} & =  \exp[\beta^{[\sigma]}_{0}+\beta^{[\sigma]}_{1}\log(S_{\Lowx})]
\end{split}
\end{align}
are the shape parameters of the distribution of $N$
at stress levels $S_{\Highx}$ and $S_{\Lowx}$, respectively, and
\begin{align}
\begin{split}
\label{equation:n.low.definition}
	t_{\Lowx} & =  \exp[\beta_{0}+\beta_{1}\nu(S_{\Highx};\lambda) + \Phi^{-1}(0.50)\sigma_{\Lowx}]\\
	t_{\Highx} & =  \exp[\beta_{0}+\beta_{1}\nu(S_{\Lowx};\lambda) + \Phi^{-1}(0.50)\sigma_{\Highx}]
\end{split}
\end{align}
are the medians of the distribution of $N$ at stress levels
$S_{\Highx}$ and $S_{\Lowx}$, respectively. Then the USPs are
$\log(\sigma_{\Lowx})$,
$\log(\sigma_{\Highx})$, $\lambda$, $\log(t_{\Lowx})$, and $\log(t_{\Highx})$.
This stable parameterization avoids the
high correlations between $\lambda$ and the regression coefficients in
the traditional parameterization.

\subsection{Initial values for the Box--Cox/Loglinear-\texorpdfstring{$\sigma_{N}$}{Lg} USPs}
\label{section:box-cox.initial.values.regression.parameters}
First, fit a simple regression $\log(N) = \beta_{0} + \beta_{1}\log(S)$ using
all of the data. Use the estimates of $\beta_{0}$, $\beta_{1}$ and
$\lambda= 0$ as start values for an NLS fitting of the model $\log(N) \sim \nu(S;\lambda)$,
providing initial values for $\beta_{0}$, $\beta_{1}$, and
$\lambda$. To keep this preliminary estimation simple, one
can treat right-censored observations 
to be failures at the censoring times.

The initial values for $\log(\sigma_{\Lowx})$ and
$\log(\sigma_{\Highx})$ are obtained as follows. Recall that
in fatigue \SN{} data, 
$\log(\sigma)$ tends to be larger at lower levels of stress.
To find the initial value for $\log(\sigma_{\Lowx})$ use the data at
the highest levels of stress (e.g., divide the data into two parts),
fit the single distribution and take the log of the resulting ML
estimate for $\sigma$.  Do the same using the lowest levels of
stress to find the initial value for $\log(\sigma_{\Highx})$.
Then calculate the  initial values for $\log(t_{\Lowx})$ and $\log(t_{\Highx})$
by substituting the NLS estimates of $\beta_{0}$, $\beta_{1}$, and
$\lambda$ into (\ref{equation:n.low.definition}) without taking antilogs.

\subsection{Computing the Box--Cox/Loglinear-\texorpdfstring{$\sigma_{N}$}{Lg} TPs from the USPs}
Similar to the previously described models
Box--Cox/Loglinear-$\sigma_{N}$ the TPs can be computed
from the Box--Cox/Loglinear-$\sigma_{N}$  USPs by
solving (\ref{equation:s.low.definition}) for
$\beta^{[\sigma]}_{1}$ and $\beta^{[\sigma]}_{0}$ giving:
\begin{align*}
\beta^{[\sigma]}_{1} &= \frac{\log(\sigma_{\Highx})-\log(\sigma_{\Lowx})}{\log(S_{\Lowx}) - \log(S_{\Highx})}\\[1ex]
\beta^{[\sigma]}_{0} &= \log(\sigma_{\Lowx}) - \beta^{[\sigma]}_{1}\log(S_{\Highx}).
\end{align*}
Then solve  (\ref{equation:n.low.definition}) for $\beta_{1}$ and $\beta_{0}$ giving:
\begin{align*}
\beta_{1} &= \frac{[\log(t_{\Highx})-\log(t_{\Lowx})]
  -(\sigma_{\Highx} - \Phi^{-1}(0.50)\sigma_{\Lowx})}{\nu(S_{\Lowx};\lambda) - \nu(S_{\Highx};\lambda)}\\[1ex]
\beta_{0} &=\log(t_{\Lowx}) - \beta_{1}\nu(S_{\Highx};\lambda) - \Phi^{-1}(0.50)\sigma_{\Lowx}.
\end{align*}

\subsection{ML estimates for the TPNSs based on
  the original unscaled data}
In the previous steps, ML estimates were obtained using scaled data,
as described in Step~\ref{step:scale.data} of
Section~\ref{section:steps.ml.estimation}.  Denote the
estimates from the scaled data by $\widetilde{\beta}_{0}$,
$\widetilde{\beta}_{1}$, $\widetilde{\lambda}$,
$\widetilde{\beta}^{[\sigma]}_{0}$, and
$\widetilde{\beta}^{[\sigma]}_{1}$, where the scaling values are
$S_{\max}$ and $N_{\max}$ for stress and number of cycles, respectively.
Then the Box--Cox/Loglinear-$\sigma_{N}$ \SN{} model for the scaled
data is:
\begin{align*}
  \log(\widetilde{N})&= \widetilde{\beta}_{0}+ \widetilde{\beta}_{1} \nu(\widetilde{S};\widetilde{\lambda})
                       +  \exp\left[
                       \widetilde{\beta}^{[\sigma]}_{0}+ \widetilde{\beta}^{[\sigma]}_{1} \log(\widetilde{S})
                       \right] \Phi^{-1}(0.50),
\end{align*}
where $\widetilde{S}$ and $\widetilde{N}$ are the scaled stress and
number of cycles, respectively.
In terms of the original unscaled data, the
model is:
\begin{align*}
  \log(N)&= \log(N_{\max})+\widetilde{\beta}_{0}
           + \widetilde{\beta}_{1} \nu(\widetilde{S};\widetilde{\lambda})
           +  \exp\left[
           \widetilde{\beta}^{[\sigma]}_{0}-\widetilde{\beta}^{[\sigma]}_{1} \log(S_{\max})
           + \widetilde{\beta}^{[\sigma]}_{1} \log(S)
           \right] \Phi^{-1}(0.50)\\
         &= \log(N_{\max})+\widetilde{\beta}_{0}
           -\frac{\widetilde{\beta}_{1}}{S^{\widetilde{\lambda}}_{\max}}\nu(S_{\max};\widetilde{\lambda})
           +\frac{ \widetilde{\beta}_{1}}{S^{\widetilde{\lambda}}_{\max}}\nu(S;\widetilde{\lambda})
           +\exp\left[
           \widetilde{\beta}^{[\sigma]}_{0}-\widetilde{\beta}^{[\sigma]}_{1} \log(S_{\max})
           + \widetilde{\beta}^{[\sigma]}_{1} \log(S)
           \right] \Phi^{-1}(0.50),
\end{align*}
which follows from the result:
\begin{align*}
  \nu\left(\widetilde{S};\widetilde{\lambda}\right) =
  \frac{1}{S^{\widetilde{\lambda}}_{\max}}\left[\nu(S;\widetilde{\lambda})-\nu(S_{\max};\widetilde{\lambda})\right].
\end{align*}
Therefore, the ML estimates for the
TPNSs based on the original unscaled data are:
\begin{align*}
  \betahat_{0} & = \log(N_{\max}) + \widetilde{\beta}_{0}-\frac{\widetilde{\beta}_{1}}{S^{\widetilde{\lambda}}_{\max}}\nu(S_{\max};\widetilde{\lambda})\\
  \betahat_{1} & = \frac{\widetilde{\beta}_{1}}{S^{\widetilde{\lambda}}_{\max}}\\
  \lambdahat & = \widetilde{\lambda}\\
  \betahat^{[\sigma]}_{0} & = \widetilde{\beta}^{[\sigma]}_{0} - \widetilde{\beta}^{[\sigma]}_{1}\log(S_{\max})\\[1ex]
  \betahat^{[\sigma]}_{1} & = \widetilde{\beta}^{[\sigma]}_{1}.
\end{align*}

\section{Numerical Examples}
\label{section:numerical.examples}
To illustrate the ideas and methods presented in this paper, this
section provides two contrasting examples. In the first example
(the Inconel 718 data)
there was a large number of test specimens. In contrast, the second example
(the Polynt data),
reflective of modern, more economical testing, is based on a test
with many fewer observed fractured specimens and a substantially larger
proportion of runouts (right-censored observations).
Section~\ref{section:summary.numerical.examples} provides a summary
of the estimation results for the collection of 29 benchmark \SN{}
data sets.

\subsection{ML estimation for the Inconel 718 data}
\label{section:estimation.Inconel.data}
\citet{Shen1994} describes and analyzes fatigue data obtained by
testing 246 specimens of Inconel 718, a nickel-base super alloy used
for components, such as turbine disks and fan blades, in the hot part
of an aircraft engine. The test was strain controlled. The data were
originally provided by General Electric Aircraft
Engines. The test resulted in 242 fractured units and 4 runouts.

Table~\ref{table:summary.Inconel.model.fits} gives a summary of
estimation results for the Inconel 718 data showing the ten models
with the smallest values of AIC. The differences among the top three
models is already substantial, indicating a clear preferences for
the Nishijima/Lognormal model. In terms of AIC, the
Nishijima/Loglogistic is a distant second place (the loglogistic
distribution has a shape similar to the lognormal, but with somewhat
heavier tails). Because of the large amount of data in this example, the AIC
provides a strong indication about which of the fitted models provides the best
description of the data.
\begin{table}[!tbp]
\caption{Comparison, for the Inconel 718 data, of the models with the
  smallest values of AIC}
\label{table:summary.Inconel.model.fits}
\begin{center}
\begin{tabular}{lllcrr}
\hline\hline
&\multicolumn{2}{c}{Model}\tabularnewline
\hline
Model Specified for&Relationship&Distribution&\#Parms& $-\loglike$&AIC\tabularnewline
\hline
Fatigue Strength&Nishijima&Lognormal&5&-1276.6&-2543.2\tabularnewline
Fatigue Strength&Nishijima&Loglogistic&5&-1273.3&-2536.5\tabularnewline
Fatigue Strength&Coffin--Manson Zero Elastic Slope&Lognormal&4&-1266.9&-2525.8\tabularnewline
Fatigue Strength&Coffin--Manson&Lognormal&5&-1266.9&-2523.8\tabularnewline
Fatigue Strength&Nishijima&Weibull&5&-1263.9&-2517.8\tabularnewline
Fatigue Strength&Coffin--Manson Zero Elastic Slope&Loglogistic&4&-1262.4&-2516.7\tabularnewline
Fatigue Strength&Coffin--Manson&Loglogistic&5&-1262.4&-2514.7\tabularnewline
Fatigue Strength&Nishijima&Frechet&5&-1258.3&-2506.6\tabularnewline
Fatigue Strength&Coffin--Manson Zero Elastic Slope&Frechet&4&-1255.1&-2502.3\tabularnewline
Fatigue Strength&Coffin--Manson&Frechet&5&-1255.1&-2500.3\tabularnewline
Fatigue Strength&Coffin--Manson Zero Elastic Slope&Weibull&4&-1252.9&-2497.8\tabularnewline
\hline
\end{tabular}
\end{center}
\end{table}
Table~\ref{table:Inconel.model.MLE} gives ML estimates of the TPNSs
and corresponding 95\% confidence intervals for the
Nishijima/Lognormal fatigue-strength model.
\begin{table}[!tbp]
\caption{ML estimates and Wald confidence intervals for the TPNSs of
  the Nishijima/Lognormal fatigue-strength model fit to the Inconel
  718 fatigue data}
\label{table:Inconel.model.MLE}
\begin{center}
\begin{tabular}{crrrr}
\hline\hline
&&&\multicolumn{2}{c}{95\% Confidence Interval}\tabularnewline
\multicolumn{1}{l}{Parameter}&\multicolumn{1}{c}{Estimate}&\multicolumn{1}{c}{Std Error}&\multicolumn{1}{c}{Lower}&\multicolumn{1}{c}{Upper}\tabularnewline
\hline
$A$&$ 0.418$&$0.01940$&$ 0.3820$&$ 0.458$\tabularnewline
$B$&$ 0.769$&$0.01970$&$ 0.7303$&$ 0.807$\tabularnewline
$C$&$ 0.123$&$0.04920$&$ 0.0566$&$ 0.270$\tabularnewline
$E$&$-1.127$&$0.04610$&$-1.2178$&$-1.037$\tabularnewline
$\sigma_{X}$&$ 0.095$&$0.00455$&$ 0.0867$&$ 0.105$\tabularnewline
\hline
\end{tabular}\end{center}
\end{table}

\begin{figure}
\begin{tabular}{cc}
(a) & (b) \\[-3.2ex]
\rsplidapdffiguresize{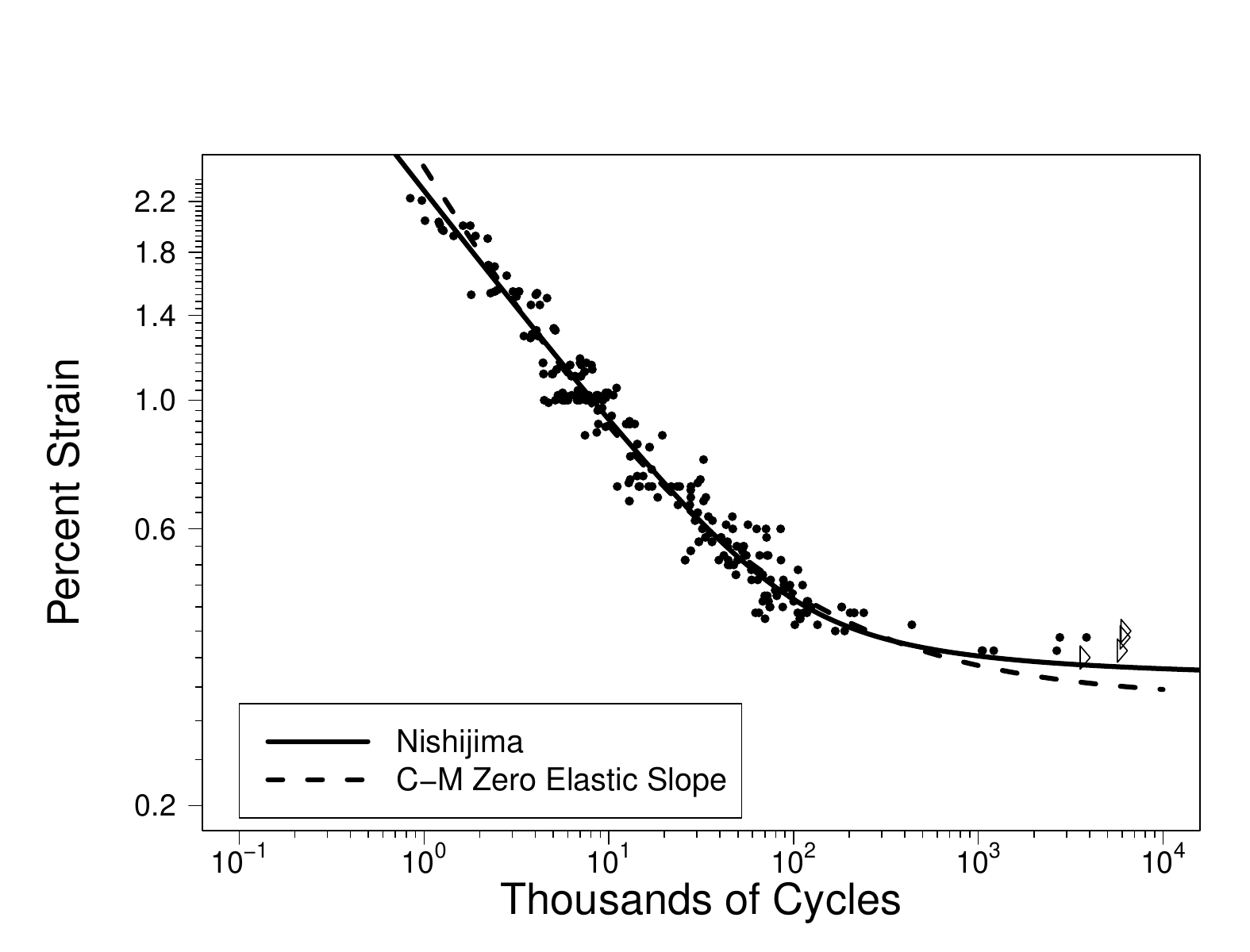}{3.25in}&
\rsplidapdffiguresize{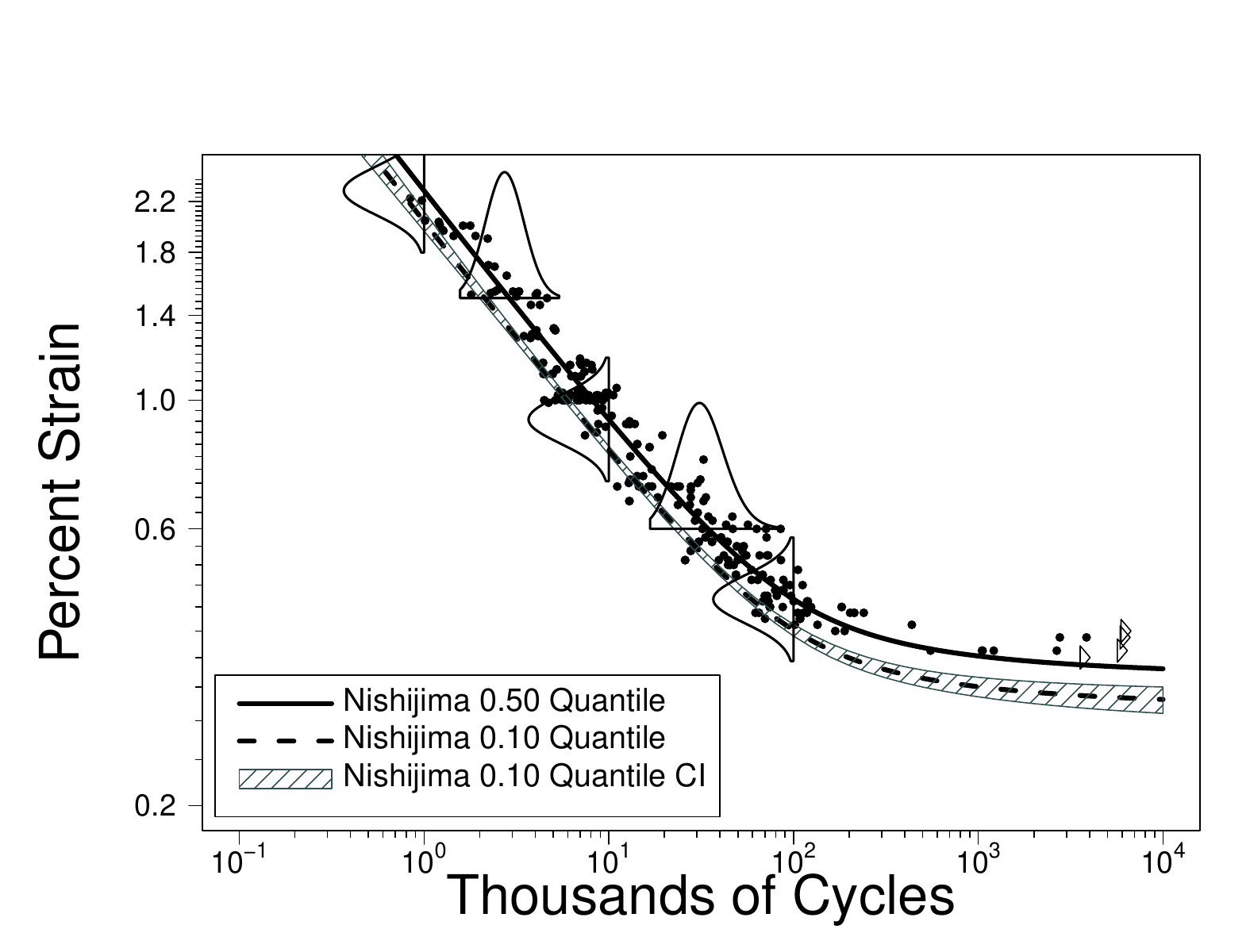}{3.25in}\\
\end{tabular}
\caption{Plot comparing the Nishijima and Coffin--Manson
  Zero-Elastic-Slope relationships (estimates of the 0.50 quantiles)
  using a lognormal distribution strength distribution with a
  constant $\sigma_{X}$ shape parameter for the Inconel 718 fatigue
  data~(a); Plot showing the ML estimates of the 0.10 and 0.50
  quantiles and 90\% likelihood-based confidence intervals for the
  0.10 quantile of the Nishijima/lognormal model~(b).}
\label{figure:Inconel.model.plots}
\end{figure}

Figure~\ref{figure:Inconel.model.plots}a 
compares the Nishijima/Lognormal and Coffin--Manson
  Zero-Elastic-Slope/Lognormal models. Focusing on the fit near the data points
  corresponding to the largest and smallest values of Thousands of
  Cycles  shows why the Nishijima/Lognormal model provides a better
  description of the data.
Figure~\ref{figure:Inconel.model.plots}b shows ML estimates of the 0.10 and 0.50
  quantiles  and the band of pointwise 90\% likelihood-based
confidence intervals for the
  0.10 quantile of the Nishijima/lognormal model. The confidence
  intervals are narrow because of the large amount of data.

\begin{figure}
\begin{tabular}{cc}
(a) & (b) \\[-3.2ex]
\rsplidapdffiguresize{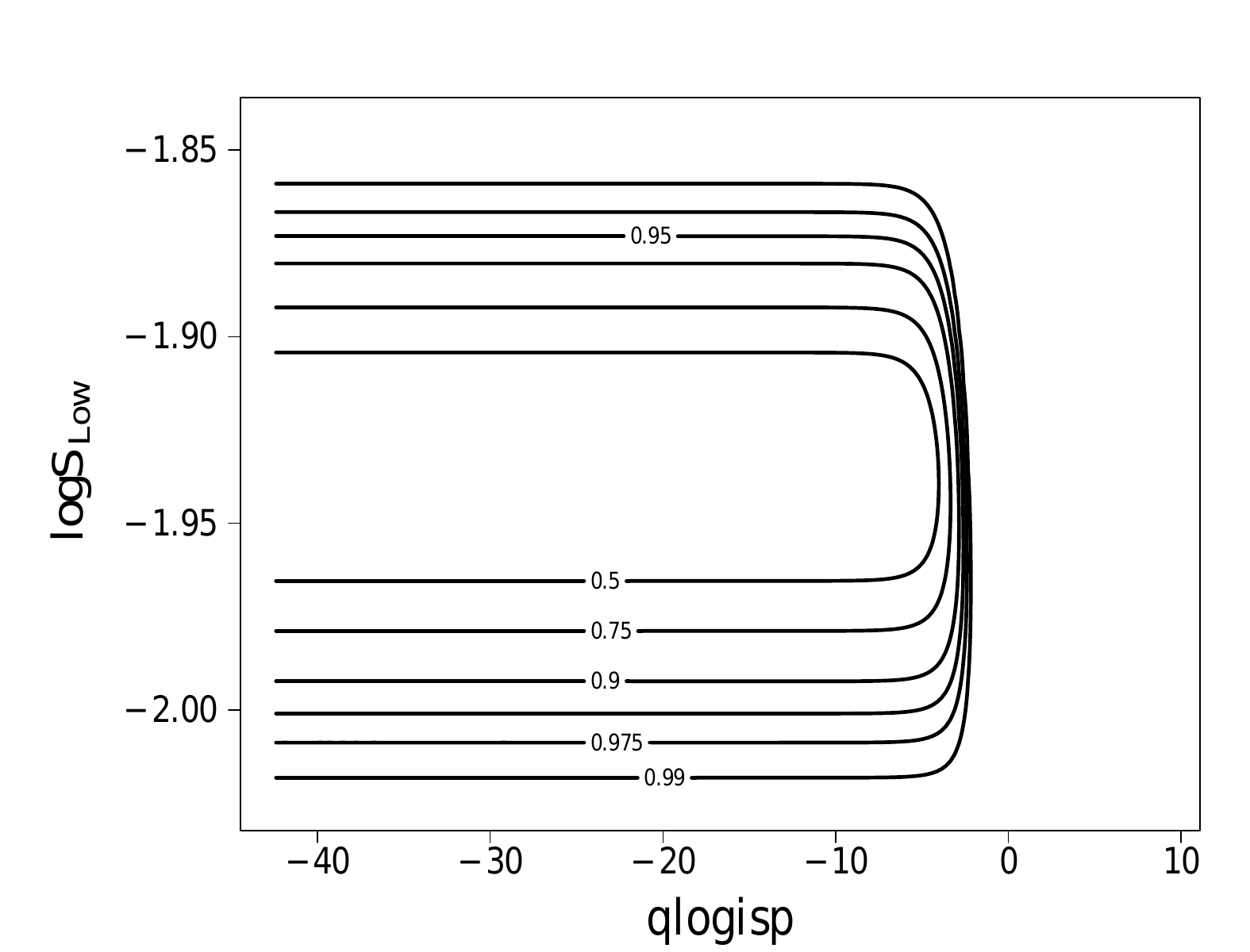}{3.25in}&
\rsplidapdffiguresize{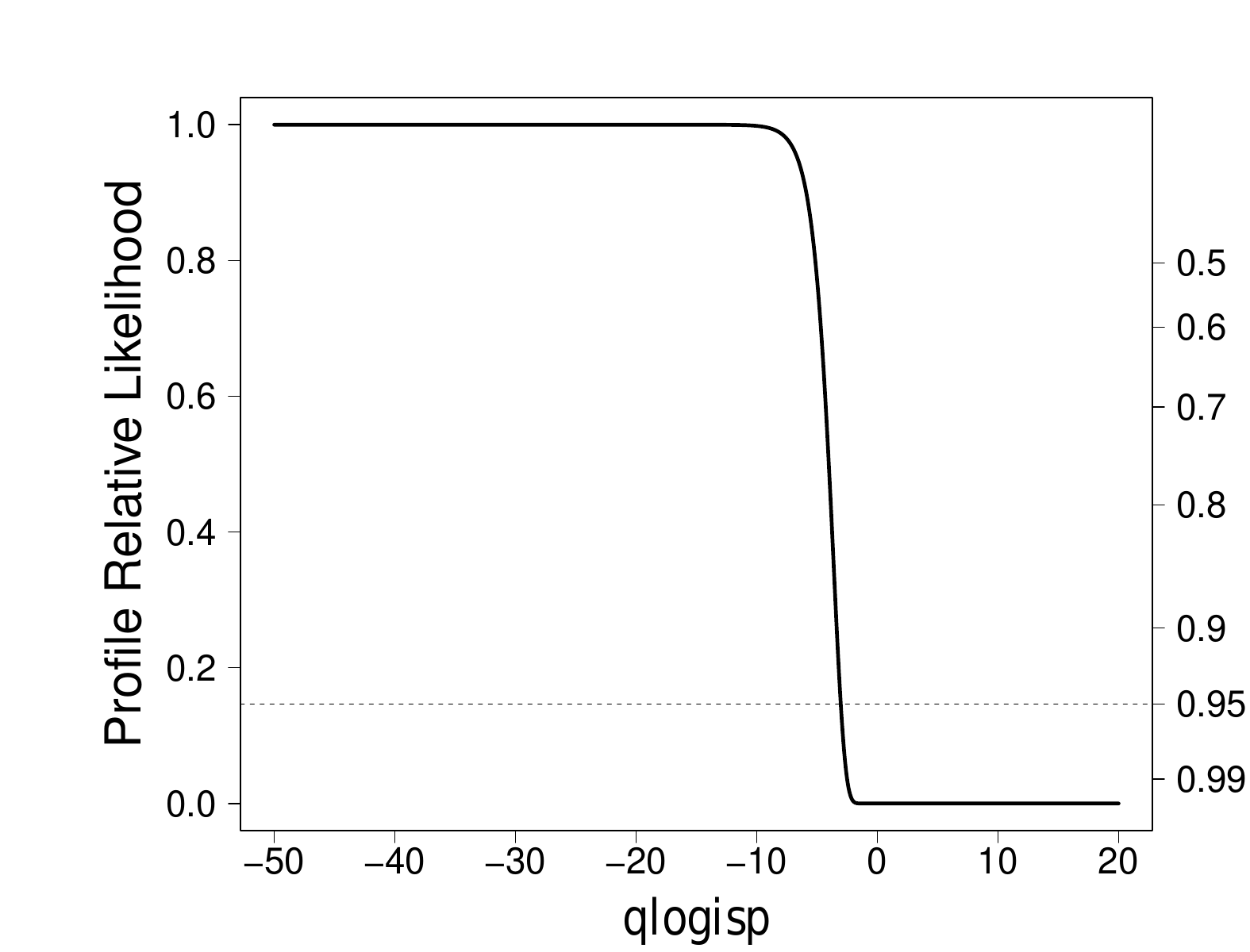}{3.25in}\\
\end{tabular}
\caption{For the Coffin--Manson model fit to the Inconel 718 data, a
  two-dimensional profile relative likelihood plot of $\qlogisp$
  and $\logSLow$~(a) and a
  relative likelihood profile plot of $\qlogisp$~(b).}
\label{figure:Inconel.profile.plots}
\end{figure}

Figure~\ref{figure:Inconel.profile.plots} illustrates, using the
Coffin--Manson model as an example, what happens when fitting an
over-parameterized model where the full model is approaching one of
the limiting models described in
Sections~\ref{section:potential.estimability.problems},
\ref{section:coffin.manson.limiting.case},~and~\ref{section:nishijima.limiting.case}.
As mentioned in Section~\ref{section:steps.ml.estimation} such plots
provide useful diagnostics if warnings about
numerical problems are encountered in
ML estimation.

Figure~\ref{figure:Inconel.profile.plots}a
is a two-dimensional relative likelihood profile plot of the
$\qlogisp$ and $\logSLow$ USPs showing a ridge along values of
$\qlogisp$ less than about
$-10$. Figure~\ref{figure:Inconel.profile.plots}a is a relative
likelihood profile plot that focuses on $\qlogisp$.  This kind of flatness in
the likelihood surface will often generate warning messages when maximizing
the log-likelihood or inverting the local estimate of the information
matrix (used for computing standard errors). The remedy is to use
the limiting Coffin--Manson Zero-Elastic-Slope model or equivalently, fix
$\qlogisp$ at a particular value along the ridge.

\subsection{ML estimation for the Polynt
composite material fatigue data}
\label{section:estimation.Polynt.data}
\citet{Tridello_etal2022} and \citet{Conte2022} describe a fatigue
test on Polynt SMC LP 2512 R33, a sheet molding compound composite
material. Dog-bone-shaped specimens were fabricated and tested in
tension-compression fatigue tests at a stress ratio $R=0.10$. Units
that survived $10^{6}$ cycles were runouts.  The first 15 specimens
were tested using a staircase test (a sequential approach that tends
to result in stress levels near the center of the fatigue-strength
distribution for a specified number of cycles). Subsequently, 20 additional
specimens were tested at higher levels of stress to provide more
information for estimating the \SN{} curve.
The test resulted in 22 fractured units and 9 runouts.
To protect proprietary information, the actual stress values
were scaled to have a maximum of 1.

Table~\ref{table:summary.Polynt.model.fits} gives a summary of
estimation results for the Polynt SMC data showing the ten models
with the smallest values of AIC. 
\begin{table}[!tbp]
\caption{Comparison, for the Polynt SMC data, of the models with the
  smallest values of AIC}
\label{table:summary.Polynt.model.fits}
\begin{center}
\begin{tabular}{lllcrr}
\hline\hline
&\multicolumn{2}{c}{Model}\tabularnewline
\hline
Model Specified for&Relationship&Distribution&\#Parms& $-\loglike$&AIC\tabularnewline
\hline
Fatigue Strength&Rectangular Hyperbola&Weibull&$4$&$-45.8$&$-83.7$\tabularnewline
Fatigue Strength&Coffin--Manson Zero Elastic Slope&Weibull&$4$&$-45.8$&$-83.5$\tabularnewline
Fatigue Life&Basquin (Inverse Power)&Weibull&$3$&$-44.5$&$-83.0$\tabularnewline
Fatigue Life&Box--Cox/Loglinear Sigma&Weibull&$5$&$-46.1$&$-82.3$\tabularnewline
Fatigue Strength&Coffin--Manson&Weibull&$5$&$-45.8$&$-81.7$\tabularnewline
Fatigue Strength&Nishijima&Weibull&$5$&$-45.8$&$-81.7$\tabularnewline
Fatigue Strength&Rectangular Hyperbola&Lognormal&$4$&$-44.8$&$-81.6$\tabularnewline
Fatigue Life&Box--Cox/Loglinear Sigma&Lognormal&$5$&$-45.7$&$-81.4$\tabularnewline
Fatigue Strength&Coffin--Manson Zero Elastic Slope&Lognormal&$4$&$-44.7$&$-81.3$\tabularnewline
Fatigue Strength&Rectangular Hyperbola&Loglogistic&$4$&$-44.5$&$-80.9$\tabularnewline
\hline
\hline
\end{tabular}
\end{center}
\end{table}
The best fitting models are the Rectangular-Hyperbola/Weibull model
(limiting model for the Nishijima/Weibull model, as described in
Section~\ref{section:nishijima.limiting.case}) and the Coffin--Manson
Zero Elastic-Slope/Weibull model (limiting model for the
Coffin--Manson/Weibull model, as described in
Section~\ref{section:coffin.manson.limiting.case}).  The AIC values
for these top two models are close to each other and the two fitted
median quantile lines (not shown here), are visually,
identical.

Interestingly, the Basquin/Weibull model (linear on log-log scales)
also has a small AIC. AIC is less able (compared with the Inconel
718 example) to discriminate among the different Polynt SMC models
because there is substantially less information in the data (only 22
fractured units and 9 runouts when compared to 242 fractured units
and 4 runouts for the Inconel 718 example).
Table~\ref{table:Polynt.model.MLE} gives ML estimates of the TPNSs and
corresponding 95\% confidence intervals for the Rectangular
Hyperbola/Weibull fatigue-strength model.
\begin{table}[!tbp]
\caption{ML estimates and Wald confidence intervals for the TPNS of
  the Rectangular Hyperbola/Weibull fatigue-strength model fit to
  the Polynt SMC  fatigue data}
\label{table:Polynt.model.MLE}
\begin{center}
\begin{tabular}{crrrr}
\hline\hline
&&&\multicolumn{2}{c}{95\% Confidence Interval}\tabularnewline
\multicolumn{1}{l}{Parameter}&\multicolumn{1}{c}{Estimate}&\multicolumn{1}{c}{Std Error}&\multicolumn{1}{c}{Lower}&\multicolumn{1}{c}{Upper}\tabularnewline
\hline
$B$&$-9.48$&8.03&$-25.21$&6.26\tabularnewline
$C$&14.2&17.27&$-19.66$&$48.05$\tabularnewline
$E$&$-1.49$&0.65&$-2.76$&$-0.22$\tabularnewline
$\sigma_{X}$&0.091&0.018&0.057&0.126\tabularnewline
\hline
\end{tabular}\end{center}
\end{table}

\begin{figure}
\begin{tabular}{cc}
(a) & (b) \\[-3.2ex]
\rsplidapdffiguresize{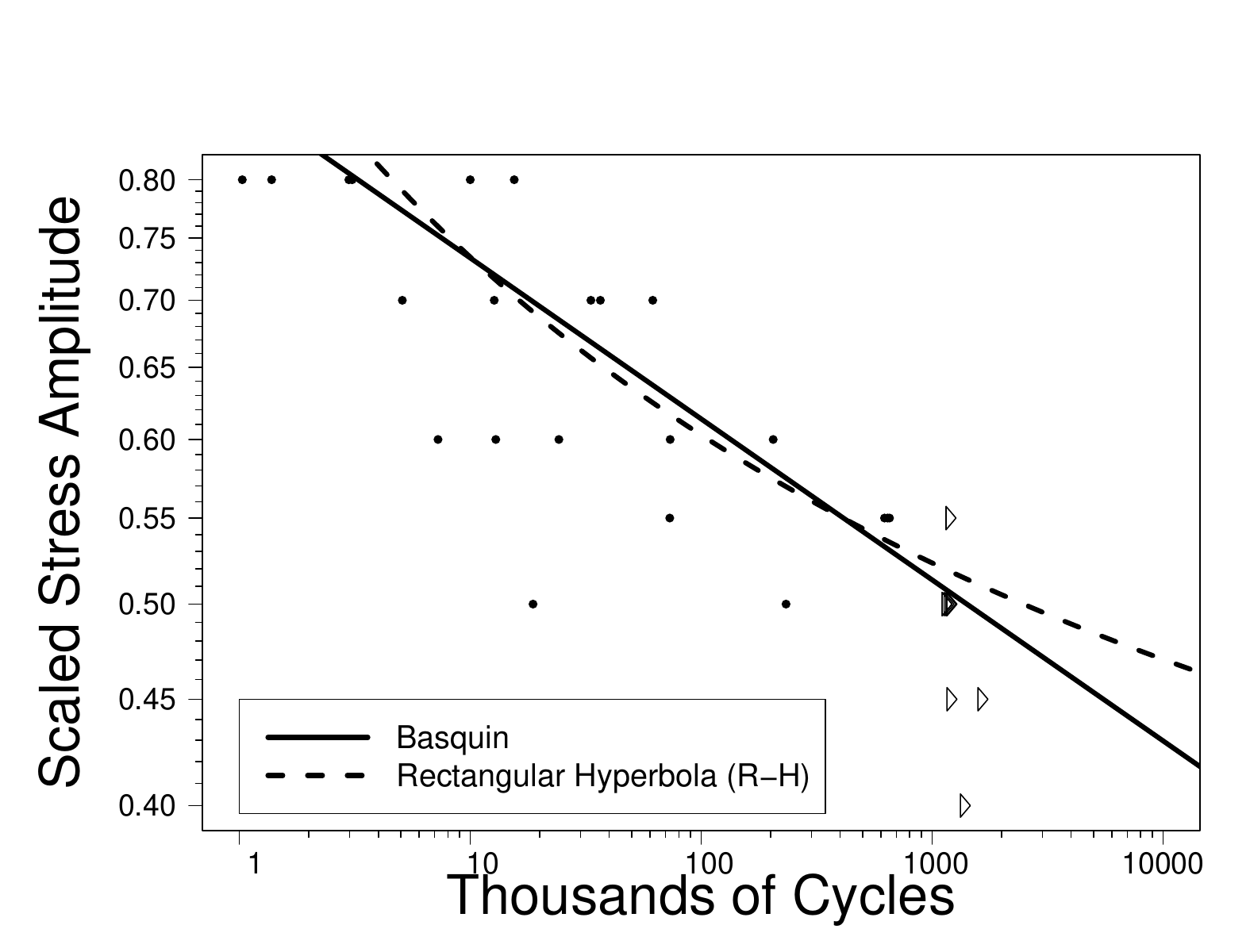}{3.25in}&
\rsplidapdffiguresize{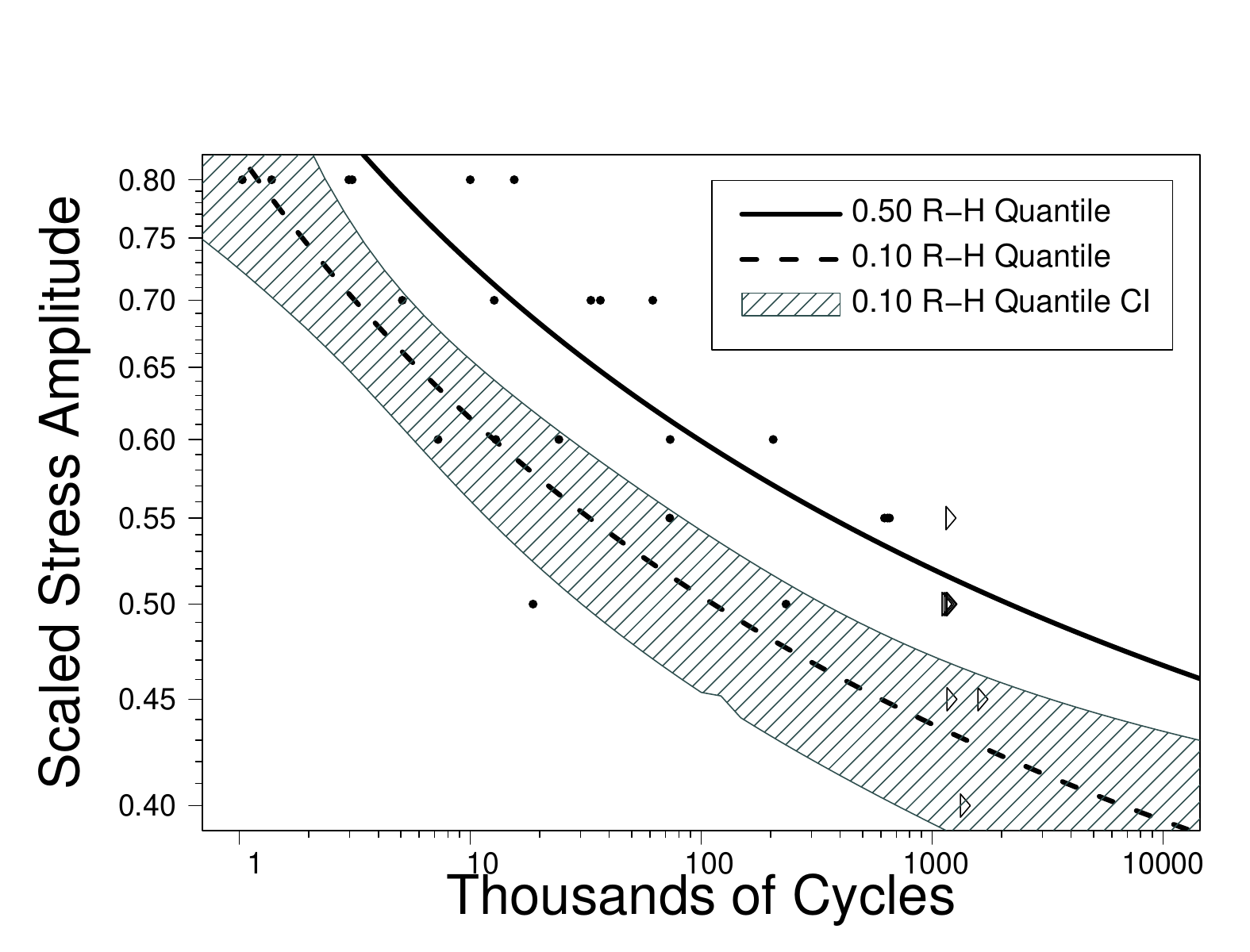}{3.25in}\\
\end{tabular}
\caption{For the
  Polynt SMC fatigue data, a plot comparing the
  Rectangular Hyperbola and Basquin
  Relationships (estimates of the 0.50 quantiles)
  using a Weibull distribution strength
  distribution with a constant  $\sigma_{X}$ shape parameter~(a); Plot
  showing the ML estimates of the 0.10 and 0.50 quantiles and
  the band of pointwise 90\%
  likelihood-based confidence intervals for the 0.10
  quantile of the Rectangular Hyperbola/Weibull model~(b).}
\label{figure:Polynt.model.comparison.plots}
\end{figure}

Figure~\ref{figure:Polynt.model.comparison.plots}a compares the
Rectangular Hyperbola/Weibull and the Basquin/Weibull models. 
The Basquin/Weibull model has a smaller likelihood and a larger AIC
(in Table~\ref{table:summary.Polynt.model.fits}) because of lack of
fit implied by the substantial number of runouts at the lower levels
of stress. Although the Basquin/Weibull is statistically consistent
with the limited data, engineering/physical knowledge about the
general nature of fatigue-life processes would suggest the
Rectangular Hyperbola
or the Coffin--Manson Zero Elastic-Slope/Weibull models better choices.

Figure~\ref{figure:Polynt.model.comparison.plots}b shows, for the
Rectangular Hyperbola/Weibull model, the ML estimates of the 0.10
and 0.50 quantiles and the band of pointwise 90\% likelihood-based
confidence intervals for the 0.10 quantile. The relative width of
the confidence bands is considerably wider than in the Inconel 718
example because there is less information in the limited Polynt SMC
data.

Next we investigate some potential numerical difficulties that can
arise when overfitting a nonlinear regression model and tools to
diagnose such difficulties. Two possible difficulties are
\begin{itemize}
\item
Regions in the parameter space where the likelihood surface is flat
or approximately flat.
\item
A likelihood surface that has more than one maximum, often a global
maximum and one or more local maxima.
\end{itemize}
The flat spots in the likelihood surface can cause warning or error
messages from both the optimization algorithm (used to maximize the
likelihood) and differentiation algorithms (used to compute the Hessian
matrix). When there is more than one maximum in the likelihood
surface, as mentioned in Section~\ref{section:steps.ml.estimation},
it is possible that the optimization algorithm will converge to a
local instead of the global optimum. As mentioned in
Section~\ref{section:steps.ml.estimation}, we have found that one-
and two-dimensional relative likelihood profiles are useful for
exploring and finding remedies for these difficulties.

\begin{figure}
\begin{tabular}{cc}
(a) & (b) \\[-3.2ex]
\rsplidapdffiguresize{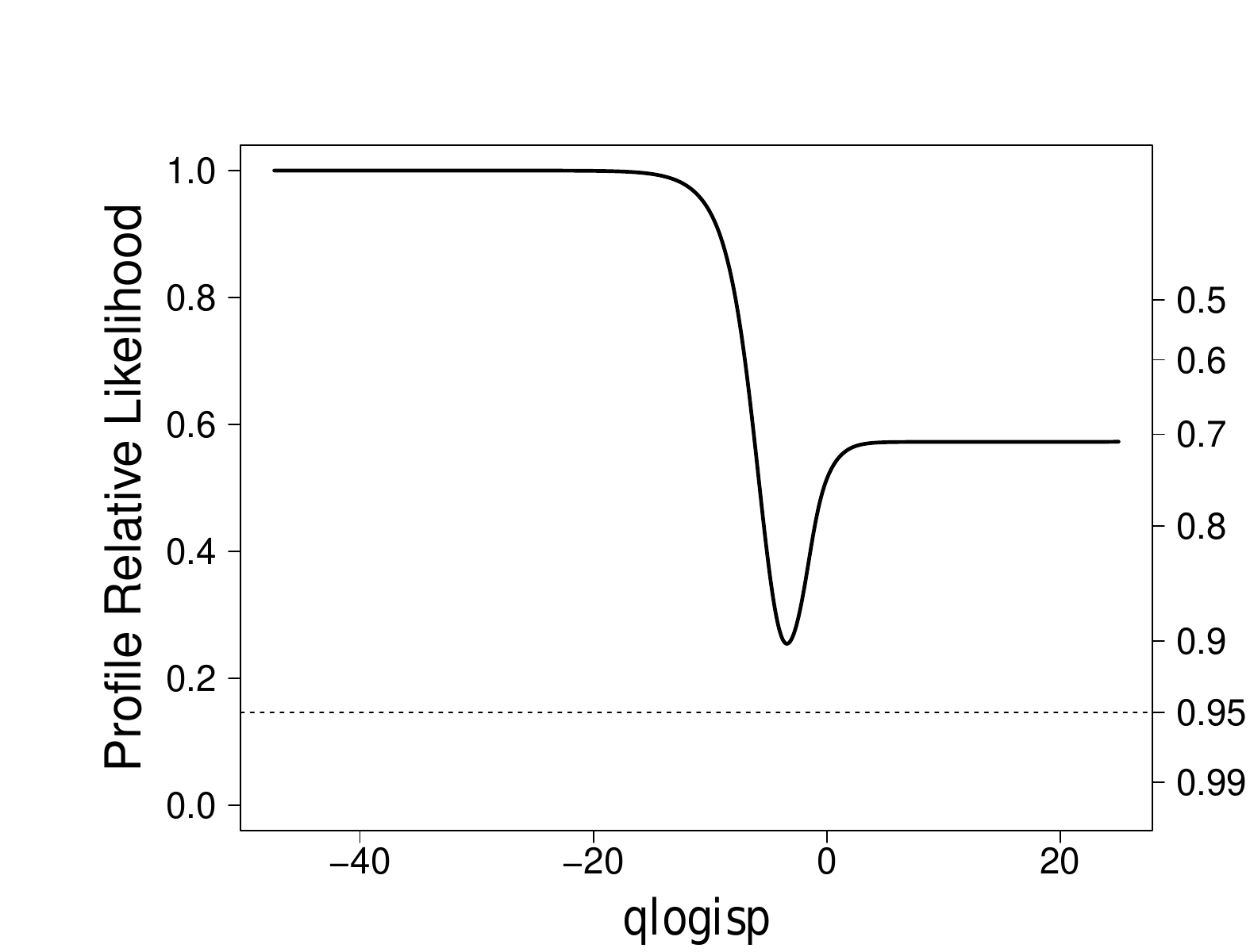}{3.25in}&
\rsplidapdffiguresize{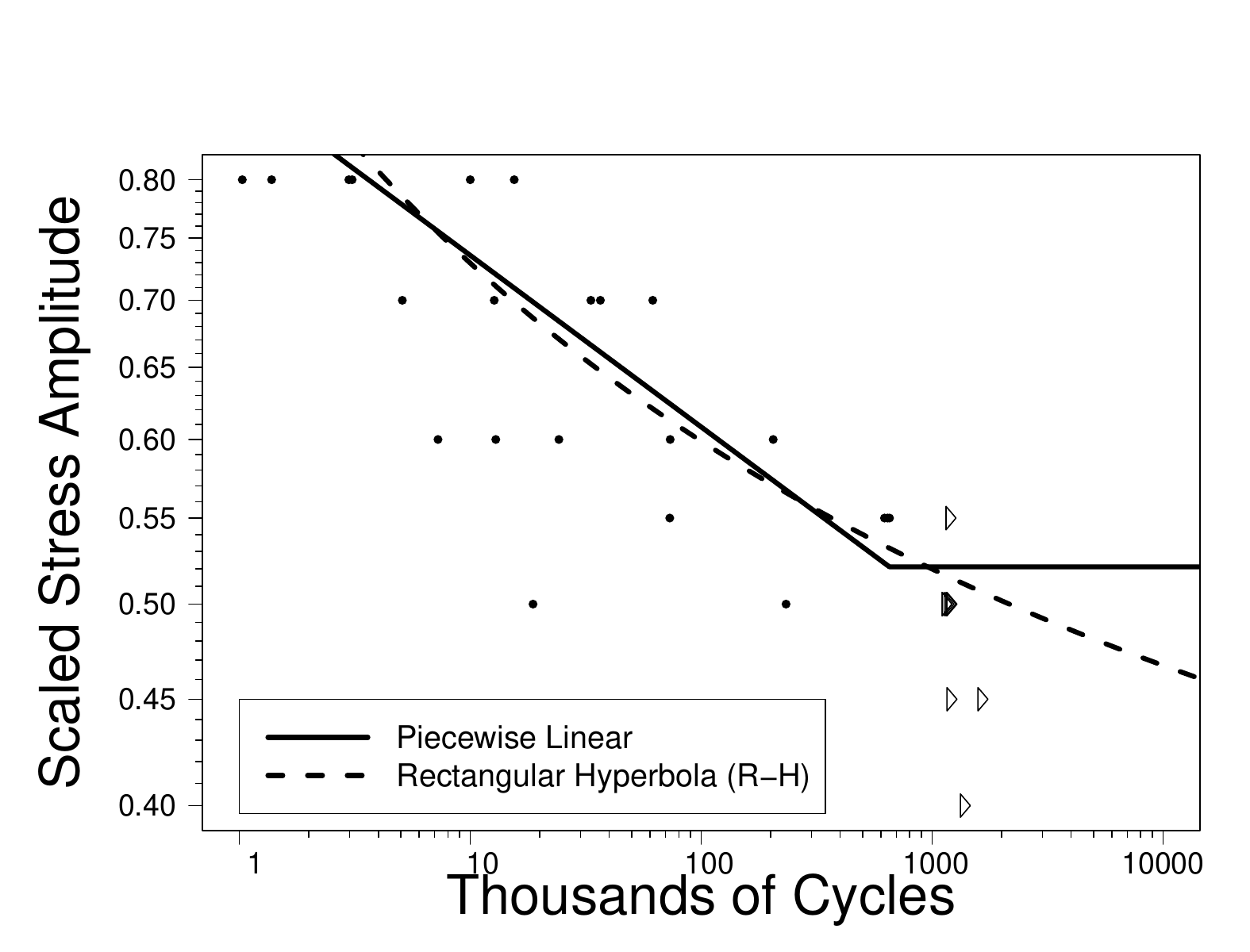}{3.25in}\\
\end{tabular}
\caption{For the
  Polynt SMC fatigue data and the Nishijima/Weibull model, profile
  relative likelihood plot of $\qlogisp$~(a) and a plot of the data
  with the two fitted models corresponding to the two flat parts of the
  profile relative likelihood plot~(b).}
\label{figure:Polynt.profile.plots}
\end{figure}

Figure~\ref{figure:Polynt.profile.plots}a is the profile relative
likelihood plot of $\qlogisp$ that arises when fitting the
Nishijima/Weibull model to the Polynt SMC fatigue data.
The two intervals of $\qlogisp$ values where this profile plot is
flat suggests two competing descriptions (models) for the data. Over
the interval from about $5$ to about $28$ the model fit corresponds
to the limiting Rectangular-Hyperbola/Weibull model described in
Section~\ref{section:nishijima.limiting.case} and illustrated by the
dashed fitted line in Figure~\ref{figure:Polynt.profile.plots}b.
Over the interval from $-\infty$ to about $-20$ the model fit
corresponds to the limiting Piecewise-Linear/Weibull model also
described in Section~\ref{section:nishijima.limiting.case} and
illustrated by the solid fitted line in
Figure~\ref{figure:Polynt.profile.plots}b. The flatness in these two
regions of the profile relative likelihood indicates that there are
multiple combinations of the parameter values that result in the
same fitted curve. The higher level of the profile relative
likelihood for the Piecewise-Linear/Weibull model suggests a
somewhat better fit, but both limiting models are consistent with the
data. As a practical matter, one could choose the model that
provides more conservative inferences. 
When $\qlogisp \rightarrow \infty$, the Nishijima model
approaches rectangular hyperbola model.
Numerical problems arise for $\qlogisp >20$ but the profile would
remain horizontally flat if the computations could be done using
unlimited precision.

\subsection{Summary of the analyses of the 
benchmark collection of fatigue-life data sets}
\label{section:summary.numerical.examples}
As mentioned in Section~\ref{section:steps.ml.estimation}, it is
useful to test newly developed estimation algorithms on a collection
benchmark data sets.
For the particular applications used in this paper,
we used a benchmark collection of 29 fatigue \SN{} data sets having different
shapes and sizes. The data sets were obtained from the statistical
and engineering literature and other sources and are available in the
online supplementary materials. One of the data sets was simulated
precisely to provide a stress test for models that would be over parameterized.

These data sets were used to exercise our computational and
statistical algorithms and to provide experience and insights into
analytical properties of the nonlinear relationships and limiting
cases for the main models described in
Sections~\ref{section:parameterization.initial.values.coffin.manson.model}--\ref{section:parameterization.initial.values.box.cox.loglinear.model}
and to fine-tune our definitions of the USPs. In
particular, for each of the 29 data sets, the
\begin{enumerate}
\item
Basquin (Inverse Power)
\item
Box--Cox/Loglinear Sigma
\item
Coffin--Manson
\item
Coffin--Manson Zero Elastic Slope
\item
Nishijima
\item
Rectangular Hyperbola
\end{enumerate}
\SN{} relationships were paired with the Fr\'{e}chet, Loglogistic,
Lognormal, and Weibull log-location-scale distributions, for a total
of 696 fitted models. The results are summarized in
Table~\ref{table:summary.fatigue.data.model.fits} which gives, for
each data set, the combination of relationship and distribution (i.e.,
the model) with the smallest AIC.

\begin{sidewaystable}[!tbp]
\caption{Model with the smallest AIC value for each of the data sets
  in the benchmark collection}
\label{table:summary.fatigue.data.model.fits}
\begin{center}
\begin{tabular}{llllcrr}
\hline\hline
\multicolumn{1}{c}{Data Set Name}&\multicolumn{1}{c}{Model Specified for}&\multicolumn{1}{c}{Relationship}&\multicolumn{1}{c}{Distribution}&\multicolumn{1}{c}{\#Parms}&\multicolumn{1}{c}{$-\loglike$}&\multicolumn{1}{c}{AIC}\tabularnewline
\hline
Aluminum2024-T4&Fatigue Strength&Nishijima&Weibull&$5$&$-1007.1$&$-2004.2$\tabularnewline
Aluminum6061-T6&Fatigue Life&Box--Cox/Loglinear Sigma&Loglogistic&$5$&$ -567.7$&$-1125.5$\tabularnewline
Aluminum6061-T6Censored&Fatigue Life&Box--Cox/Loglinear Sigma&Loglogistic&$5$&$ -485.2$&$ -960.4$\tabularnewline
AnnealedAluminumWire&Fatigue Life&Box--Cox/Loglinear Sigma&Weibull&$5$&$ -626.9$&$-1243.7$\tabularnewline
AnnealedCopperWire&Fatigue Life&Box--Cox/Loglinear Sigma&Loglogistic&$5$&$ -578.3$&$-1146.5$\tabularnewline
AnnealedIronWire&Fatigue Strength&Nishijima&Loglogistic&$5$&$ -655.7$&$-1301.3$\tabularnewline
Basquin.sim30&Fatigue Life&Basquin (Inverse Power)&Lognormal&$3$&$  -40.8$&$  -75.5$\tabularnewline
BKfatigue02&Fatigue Strength&Nishijima&Weibull&$5$&$ -167.9$&$ -325.8$\tabularnewline
C35Steel&Fatigue Strength&Coffin--Manson Zero Elastic Slope&Lognormal&$4$&$ -772.0$&$-1536.1$\tabularnewline
CeramicBearing02&Fatigue Life&Basquin (Inverse Power)&Weibull&$3$&$  -90.5$&$ -174.9$\tabularnewline
Concrete&Fatigue Strength&Nishijima&Lognormal&$5$&$ -514.4$&$-1018.8$\tabularnewline
Concrete2&Fatigue Strength&Nishijima&Lognormal&$5$&$ -361.8$&$ -713.6$\tabularnewline
GAS7C3-5GM-TA&Fatigue Strength&Coffin--Manson Zero Elastic Slope&Loglogistic&$4$&$  -88.5$&$ -169.1$\tabularnewline
Inconel718&Fatigue Strength&Nishijima&Lognormal&$5$&$-1276.6$&$-2543.2$\tabularnewline
Inconel718LowStrain&Fatigue Strength&Nishijima&Loglogistic&$5$&$ -399.8$&$ -789.6$\tabularnewline
LaminatePanel&Fatigue Strength&Rectangular Hyperbola&Lognormal&$4$&$ -259.2$&$ -510.3$\tabularnewline
MagnesiumAlloyAZ61&Fatigue Life&Box--Cox/Loglinear Sigma&Weibull&$5$&$  -78.8$&$ -147.5$\tabularnewline
NickelWire&Fatigue Strength&Nishijima&Weibull&$5$&$ -739.8$&$-1469.5$\tabularnewline
Nitinol02&Fatigue Strength&Coffin--Manson&Lognormal&$5$&$ -379.9$&$ -749.8$\tabularnewline
Nitinol03&Fatigue Strength&Coffin--Manson&Lognormal&$5$&$ -497.2$&$ -984.3$\tabularnewline
PolyintSMC&Fatigue Strength&Rectangular Hyperbola&Weibull&$4$&$  -45.8$&$  -83.7$\tabularnewline
SteelCruciformShaped&Fatigue Life&Box--Cox/Loglinear Sigma&Lognormal&$5$&$  -80.1$&$ -150.2$\tabularnewline
SteelS355J2&Fatigue Strength&Coffin--Manson Zero Elastic Slope&Lognormal&$4$&$ -171.2$&$ -334.3$\tabularnewline
SteelWire&Fatigue Strength&Nishijima&Loglogistic&$5$&$ -456.1$&$ -902.3$\tabularnewline
SuperAlloy&Fatigue Strength&Coffin--Manson&Weibull&$5$&$  -24.8$&$  -39.5$\tabularnewline
TBC600Y980T&Fatigue Strength&Coffin--Manson Zero Elastic Slope&Lognormal&$4$&$  -32.1$&$  -56.2$\tabularnewline
Ti64-350F-Rm1Corrected&Fatigue Strength&Rectangular Hyperbola&Lognormal&$4$&$ -210.1$&$ -412.2$\tabularnewline
Titanium02&Fatigue Life&Box--Cox/Loglinear Sigma&Frechet&$5$&$ -202.4$&$ -394.8$\tabularnewline
Triax&Fatigue Strength&Nishijima&Frechet&$5$&$ -201.6$&$ -393.3$\tabularnewline
UltraCleanSteel&Fatigue Strength&Rectangular Hyperbola&Lognormal&$4$&$  -25.4$&$  -42.7$\tabularnewline
\hline
\end{tabular}
\end{center}
\end{sidewaystable}

For many of the data sets, several models had values of AIC that
were very close to the the value for the model with the smallest
AIC. In these cases, these fitted models agree well with the best
model within the range of the data. In some cases there was an
indication of possible convergence failure. For example, in some
cases the profile likelihood for a parameter will be approximately
flat at a level that is well above 0 (but less than one) as the
parameter value approaches $-\infty$ or $\infty$, indicating that
the fitted model is not far from one of the limiting models.  These
occurrences were often for data sets where a limiting model has a
higher AIC because the data does not contain enough information to
estimate, adequately, all of the parameters of the full model.

\section{Concluding Remarks and Areas for Future Research}
\label{section:concluding.remarks}
Nonlinear regression presents challenges for users. Especially when
the same group of models will be used frequently for analyses with
many different data sets, it is useful to study the analytical
properties of the models and design appropriate robust numerical
methods for those models. Based on our experiences we have outlined
a general procedure for doing this and have provided the details for
three of the recent models for which we have developed such
algorithms. The ideas here readily extend to applications with more
than one explanatory variable and to hierarchical models although
there are likely to be some challenges in working out the details.

\section{Acknowledgments}
Luis A. Escobar provided helpful comments on an earlier version of this paper.
 Professor Andrea Tridello  provided a copy of the Polynt SMC data
 that we used as an example in Section~\ref{section:estimation.Polynt.data}.
\clearpage
{
\bibliographystyle{chicago}
\addcontentsline{toc}{section}{\protect\numberline{}References}
\bibliography{ms}
}
\end{document}